\documentclass[12pt,preprint]{aastex}
\usepackage{graphicx}
\usepackage{emulateapj5}           

\slugcomment{Accepted by ApJ on April 4th, 2005}

\shorttitle{NH$_3$ and CH$_3$OH emission toward NGC6334\,I \& I(N)}
\shortauthors{Beuther et al.}

\begin{document}
\newcommand{\methone}{CH$_3$OH($3_{2,1}-3_{1,2}$)E}
\newcommand{\methtwo}{CH$_3$OH($4_{2,2}-4_{1,3}$)E}
\newcommand{\meththree}{CH$_3$OH($2_{2,0}-2_{1,1}$)E}


\title{High-spatial resolution observations of NH$_3$ and CH$_3$OH towards the massive twin cores NGC6334\,I \& I(N)}


\author{H. Beuther, S. Thorwirth, Q. Zhang, T.R. Hunter, S.T. Megeath}

\affil{Harvard-Smithsonian Center for Astrophysics, 60 Garden Street, Cambridge, MA 02138, USA}

\email{hbeuther@cfa.harvard.edu}

\author{A.J. Walsh}
\affil{Department of Astrophysics and Optics, University of New South Wales, NSW 2052, Australia}


\author{K.M. Menten}
\affil{Max-Planck-Institut f\"ur Radioastronomie, Auf dem H\"ugel 69, 53121 Bonn, Germany}




\begin{abstract} Molecular line observations of NH$_3$ (J,K)=(1,1),
(2,2) and CH$_3$OH at 24.93\,GHz taken with the Australian Telescope
Compact Array (ATCA) toward the massive twin cores NGC6334\,I \& I(N)
reveal significant variations in the line emission between the two
massive cores. The UCH{\sc ii} region/hot core NGC6334\,I exhibits
strong thermal NH$_3$ and CH$_3$OH emission adjacent to the UCH{\sc
ii} region and coincident with two mm continuum peaks observed by
Hunter et al.\,(in prep.). In contrast, we find neither compact NH$_3$
nor thermal CH$_3$OH line emission toward NGC6334\,I(N). There, the
NH$_3$ emission is distributed over a broad region ($>1'$) without a
clear peak, and we find Class I CH$_3$OH maser emission with peak
brightness temperatures up to 7000\,K. The maser emission peaks appear
to be spatially associated with the interfaces between the molecular
outflows and the ambient dense gas. Peak NH$_3$(1,1) line brightness
temperatures $\geq 70$\,K in both regions indicate gas temperatures of
the same order. NH$_3$ emission is also detected toward the outflow in
NGC6334\,I resulting in an estimated rotational temperature of
$T_{\rm{rot}}\sim 19$\,K.  Furthermore, we observe CH$_3$OH and NH$_3$
absorption toward the UCH{\sc ii} region, the velocity structure is
consistent with expanding molecular gas around the UCH{\sc ii}
region. Thermal and kinematic effects possibly imposed from the
UCH{\sc ii} region on the molecular core are also discussed.
\end{abstract}


\section{Introduction}

The prominent molecular cloud/H{\sc ii} region complex NGC6334 at a
distance of 1.7\,kpc is composed of a series of luminous star forming
regions at various stages of evolution \citep{neckel1978,straw1989}.
During the initial radio and infrared studies completed over twenty
years ago, a confusing family of nomenclature was introduced for this
region.  Letters A-F refer to VLA 6cm continuum sources
\citep{rodriguez1982}, while Roman numerals I through VI refer to
far-infrared sources \citep{mcbreen1979,loughran1986}.  While each
center of activity contains CO hot spots and various masers, the
youngest region appears to be radio source ``F'' (which corresponds to
infrared source ``I'' and IRAS source 17175--3544) located at the
northeastern end of the complex.  Even the earliest far-infrared maps
\citep{gezari1982} resolved the emission of source I into two cores:
the sources I and I(N), which are separated by $\sim90''$.  With
the cold dust emission, only a few red infrared sources and weak cm
continuum emission (e.g.,
\citealt{gezari1982,tapia1996,sandell2000,carral2002}), Source I(N) is
considered to be the least evolved object in the region.

\subsection{NGC6334\,I} 

A well-known, cometary compact H{\sc ii} region lies at the heart of
source I (e.g., \citealt{carral1997}) and contains associated OH
masers \citep{brooks2001,gaume1987}, Class II CH$_3$OH masers
\citep{ellingsen1996} and H$_2$O masers \citep{moran1980,foster1989}.
Recent mid-infrared imaging with OSCIR at the Keck II telescope has
identified a candidate star (IRS1E) for the excitation source of the
compact H{\sc ii} region \citep{debuizer2002}. Single dish bolometer
maps from JCMT reveal a compact ($<20''$) strong (sub)mm continuum
source with an estimated dust temperature of 100\,K
\citep{sandell2000}.  A SEST line-survey \citep{thorwirth2003} shows 
rich molecular line emission in the mm wavelength range similar to
prototypical high-mass star-forming regions such as Orion-KL.
Comparable results are obtained in a mini line survey near 345\,GHz
\citep{mccutcheon2000}.  A CO(2--1) outflow with a velocity range of
$\sim 150$\,km\,s$^{-1}$ emanates from NGC6334\,I
\citep{bachiller1990} and a similar outflow is also seen in the
H76$\alpha$ and H92$\alpha$ recombination lines
\citep{depree1995}. Several knots of shock-excited H$_2$ emission have
been detected near the ends of this outflow
\citep{davis1995,persi1996,megeath1999} along with NH$_3$ (3,3) masers
\citep{kraemer1995}. The driving source(s) for the outflow were
claimed to be either the powering source of the UCH{\sc ii} region, or
an adjacent object. One candidate is the cold $20\mu$m source I:KDJ2
located $6.5''$ to the north-west of the UCH{\sc ii} region (as
resolved by MIRAC2 at IRTF), which is not seen at shorter wavelengths
\citep{kraemer1999a}.  Alternatively, the powering source of the
UCH{\sc ii} region and I:KDJ2 may both drive independent outflows.

\subsection{NGC6334\,I(N)} 

Similar to source I, the JCMT bolometer maps show I(N) to be a strong
source of (sub)mm continuum emission \citep{sandell2000}. The
estimated dust temperature is lower around 30\,K. While source I is
stronger than I(N) in CS (5--4) \citep{kraemer1999b} and in CS (7--6),
the HC$_3$N (15--14) line is much stronger in I(N) than in I
\citep{megeath1999,sollins2004c}. Further differences can be seen in
the 230 GHz line survey where the CH$_3$OH bands (at 240-242\,GHz and
250-252\,GHz) are fainter in I(N) than I \citep{thorwirth2003},
probably indicating a younger age for this region. Similar conclusions
were drawn by \citet{mccutcheon2000} based on their smaller line
survey and spectral line mapping of the two regions. Indeed, in
striking contrast to I, the I(N) region shows no mid-infrared emission
and only a few red infrared sources \citep{tapia1996}.  Traditionally,
I(N) has also been known as lacking cm continuum emission; however,
two faint ($\sim 0.3$~mJy), compact cm sources have recently been
detected with the VLA \citep{carral2002}.  One object sits only $7''$
from the (sub)millimeter continuum peak, and is consistent with the
ionizing radiation from a B2 ZAMS star. This cm peak is associated
with Class II CH$_3$OH maser emission
\citep{ellingsen1996}. Furthermore, Class I CH$_3$OH maser emission
has been mapped at 44\,GHz and shows emission peaks distributed over
approximately $30''$ \citep{kogan1998}. A bipolar outflow has also
been mapped in the SiO(2--1) and (5--4) transitions with the SEST in
south-east north-west direction approximately coincident with the
distribution of the Class I CH$_3$OH maser at 44\,GHz
\citep{megeath1999}.  These authors also find independent evidence for
outflow activity in H$_2$ clumps, which are present but less
pronounced than in source I.

\subsection{Concerted ATCA and SMA observations}

Because NGC6334\,I \& I(N) are close to each other~-- thus sharing a
similar large-scale environment~-- but in different evolutionary
stages, they are the ideal target sources to characterize the
evolutionary differences in physical and chemical properties of
massive star-forming regions. Both sources have been thoroughly
investigated at near/mid-infrared wavelength and at lower spatial
resolution in the radio and mm regime (with some already mentioned
exceptions). However, high-spatial resolution molecular line and
continuum studies from short cm to submm wavelength have been lacking
so far because of its location in the southern hemisphere. This
situation has changed over the last couple of years with the advent of
the 12\,mm and 3\,mm systems at the Australian Telescope Compact Array
(ATCA) and the dedication of the Submillimeter Array (SMA). Therefore,
we started a combined effort to investigate the molecular line and
continuum emission from 12\,mm to the submm regime employing both
instruments. Here, we present the NH$_3$(1,1), (2,2) and the CH$_3$OH
(near 25\,GHz) results from the ATCA, whereas a separate paper
outlines the first SMA results toward these sources at 217\,GHz
(Hunter et al., in prep.).

The SMA data show three mm continuum sources toward NGC6334\,I, two of
them adjacent to the UCH{\sc ii} region and one approximately
coincident with the cm continuum peak (Hunter et al.~in prep.). Toward
NGC6334\,I(N), they detect another mm continuum peak (a few arcsecond
offset from the cm-peak) with a little bit of more extended
emission. These mm peak positions will be compared with our NH$_3$ and
CH$_3$OH emission throughout this paper. Furthermore, Hunter et
al. (in prep.) detect very collimated SiO(5--4) outflows in NGC6334\,I
\& I(N), and the outflow orientation in both sources is approximately
the same in the north-east south-west direction. It is important to
mention that the new outflow in NGC6334\,I(N) is approximately
perpendicular to the previously observed larger-scale outflow
\citep{megeath1999}.

\section{Observations}

The sources NGC6334\,I \& I(N) were observed on July 6th and 7th of
2004 with the Australian Telescope Compact Array (ATCA) in the compact
750D configuration with 5 antennas and a baseline range between 31 and
719\,m.  Antenna 6 was also included, resulting in a maximum projected
baseline of $\sim 4.3$\,km. The phase reference centers are given in
Table \ref{center}. The observed lines are the NH$_3$(1,1) and (2,2)
inversion lines, with the frequencies of the main hyperfine components
at 23.695\,GHz and 23.723\,GHz, and one spectral setup at 24.932\,GHz
covering three CH$_3$OH lines at 24.9287\,GHz (\methone), 24.9335\,GHz
(\methtwo) and 24.9344\,GHz (\meththree). In addition, we observed the
cm continuum emission with one spectral unit covering a bandwidth of
128\,MHz centered at 22.08\,GHz. A good uv-coverage was obtained
through regular switching between both sources and the three spectral
setups. The spectral resolution of the observations was 16\,kHz
corresponding to a velocity resolution of $\sim
0.2$\,km\,s$^{-1}$. The primary beam of the ATCA at the given
frequency is $\sim 130''$. The data were reduced with the MIRIAD
package.  Applying a robust weighting of 1 (closer to natural than
uniform weighting, thus stressing the shorter baselines) the
synthesized beam is $2.8''\times 2.2''$. We smoothed the data to
0.5\,km\,s$^{-1}$ channels, and the rms per 0.5\,km\,s$^{-1}$ is $\sim
6$\,mJy. The rms of the 22\,GHz continuum and some of the
velocity-integrated line maps are partly slightly higher, mainly
because of more pronounced side-lobe effects due to the strong
emission features. The $3\sigma$ values are given in each Figure
separately .

\section{Observational Results}

The spatial distribution of the molecular gas is significantly
different between NGC6334\,I and NGC6334\,I(N) (Figures
\ref{ngc6334i_images} \& \ref{ngc6334in_images}). Toward NGC6334\,I, we
find strong thermal, compact NH$_3$ and CH$_3$OH emission in close
proximity to the UCH{\sc ii} region. In contrast, the NH$_3$ emission
toward NGC6334\,I(N) exhibits very extended emission without a compact
component, and we find no thermal CH$_3$OH feature but only strong
maser emission.

\subsection{NGC6334\,I} 

\begin{figure}
\includegraphics[angle=-90,width=16.6cm]{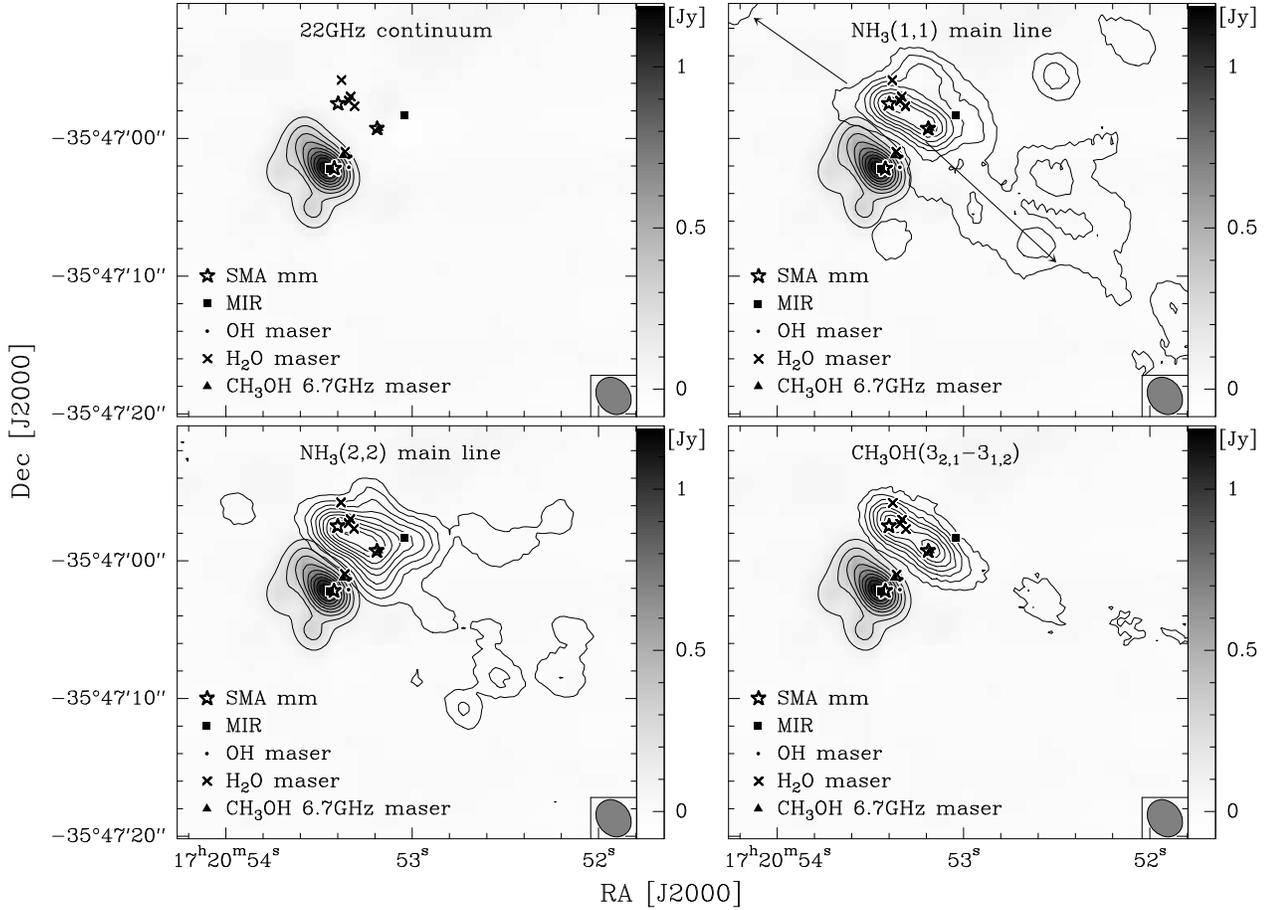}
\caption{Centimeter continuum (grey-scale and contours in the top left
image, grey-scale in the other panels), integrated main hyperfine
component NH$_3$ (1,1) (top right), (2,2) (bottom left) and
\methone~integrated line emission toward NGC6334\,I. The arrows in the
top-right panel outline the direction of the red and blue outflow
lobes (which are apparently a little bit bent, Hunter et al.~in
prep.). The contour levels are from 10 to 90\% (step 10\%) from the
peak emission of 1197\,mJy\,beam$^{-1}$ for the cm continuum image ($3\sigma
\sim 42$\,mJy\,beam$^{-1}$), and in $3\sigma$ steps for the line emission. The
approximate $3\sigma$ values for the NH$_3$(1,1), (2,2) and CH$_3$OH
lines are 21\,mJy\,beam$^{-1}$, 12\,mJy\,beam$^{-1}$ and 10\,mJy\,beam$^{-1}$,
respectively. The symbols and spectral lines are labeled in each
panel. The synthesized beams are shown at the bottom-right of each
panel as well.}
\label{ngc6334i_images}
\end{figure}

The NH$_3$ and CH$_3$OH emission from NGC6334\,I shows two emission
peaks a few arc-seconds offset from the UCH{\sc ii} region
(Fig.~\ref{ngc6334i_images}). The low-level molecular emission
terminates at the edge of the UCH{\sc ii} region, thus we have an
example of an observed edge-on orientation between an UCH{\sc ii}
region next to a hot core. The two molecular line peaks are coincident
with two of the three mm continuum peaks observed by Hunter et al.~(in
prep.) with the SMA. Interestingly, the molecular line peaks are of
the same intensity whereas the north-eastern mm-peak has about 2.5
times the continuum flux than the south-western mm-peak. Another
difference between the two peak positions is that the north-eastern peak
is associated with H$_2$O maser emission \citep{moran1980,foster1989}
whereas the south-western peak exhibits Class II CH$_3$OH maser
emission \citep{ellingsen1996}. More CH$_3$OH, H$_2$O and OH maser
emission is found at the interface between the molecular core and the
UCH{\sc ii} region. Furthermore, we detect extended
NH$_3$(1,1) emission south-west of the two peak positions. This
extended emission is significantly weaker in the NH$_3$(2,2) line and
is barely detected in CH$_3$OH. It is in the same direction as the
molecular outflow observed in CO by \citet{bachiller1990} and more
recently in SiO by Hunter et al.~(in prep.).

The spatial distribution of the NH$_3$ gas we have observed varies
considerably from the previously reported VLA map by
\citet{jackson1988}. With a spatial resolution of $5.2''\times 7.7''$,
they detected more extended emission and their peak intensity position
is $\sim 5''$ south of ours. Tapering our data to the same lower
spatial resolution, we recover most of their extended structure, but
our peak positions remains coincident with the two peaks shown in
Figure \ref{ngc6334i_images}. Due to the large number of antennas at
the VLA (27 versus 6 at the ATCA), their uv-coverage should be good
and allow a reasonable recovery of the large-scale structure, but the
offset of their peak position remains a puzzle. However, our data are
less sensitive to large-scale emission and therefore trace the
small-scale cores far better. Additionally, the exact coincidence of
our NH$_3$ and CH$_3$OH peak positions with the mm peaks reported by
Hunter et al.~(in prep.) give additional confidence to the small-scale
gas core distribution presented in Figure
\ref{ngc6334i_images}. Furthermore, considering the multiple sources
we find in NGC6334\,I and the now known outflow in the region
\citep{bachiller1990}, the disk-interpretation of the NH$_3$ data by
\citet{jackson1988} does not hold anymore, and it is more likely that
their observed large-scale emission is somehow associated with the
outflow as well.

\subsection{NGC6334\,I(N)} 

We do not detect any compact thermal NH$_3$ or CH$_3$OH emission
toward NGC6334\,I(N) (Fig.~\ref{ngc6334in_images}). However, the
extended NH$_3$ emission has about the same peak intensities as
NGC6334\,I but is distributed over a large area ($>1'$) without a
distinct peak position. The mm peak detected by Hunter et al.~(in
prep.) is located approximately at the center of the large-scale
NH$_3$ core. The NH$_3$(1,1) and (2,2) emission has a similar spatial
distribution as the NH$_3$(3,3) emission previously reported by
\citet{kraemer1999b}.

\begin{figure}
\includegraphics[angle=-90,width=8.5cm]{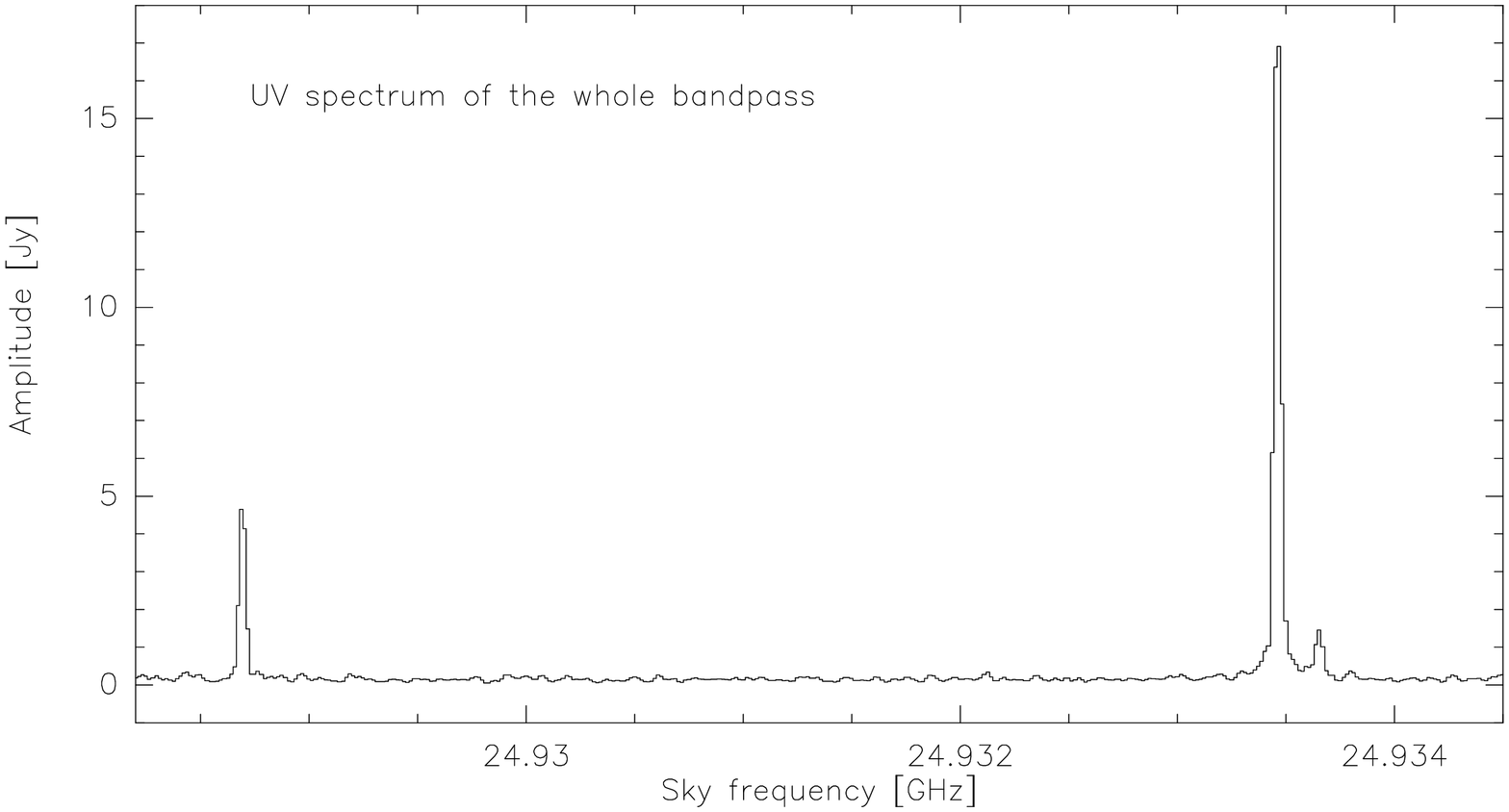}
\includegraphics[angle=-90,width=8.5cm]{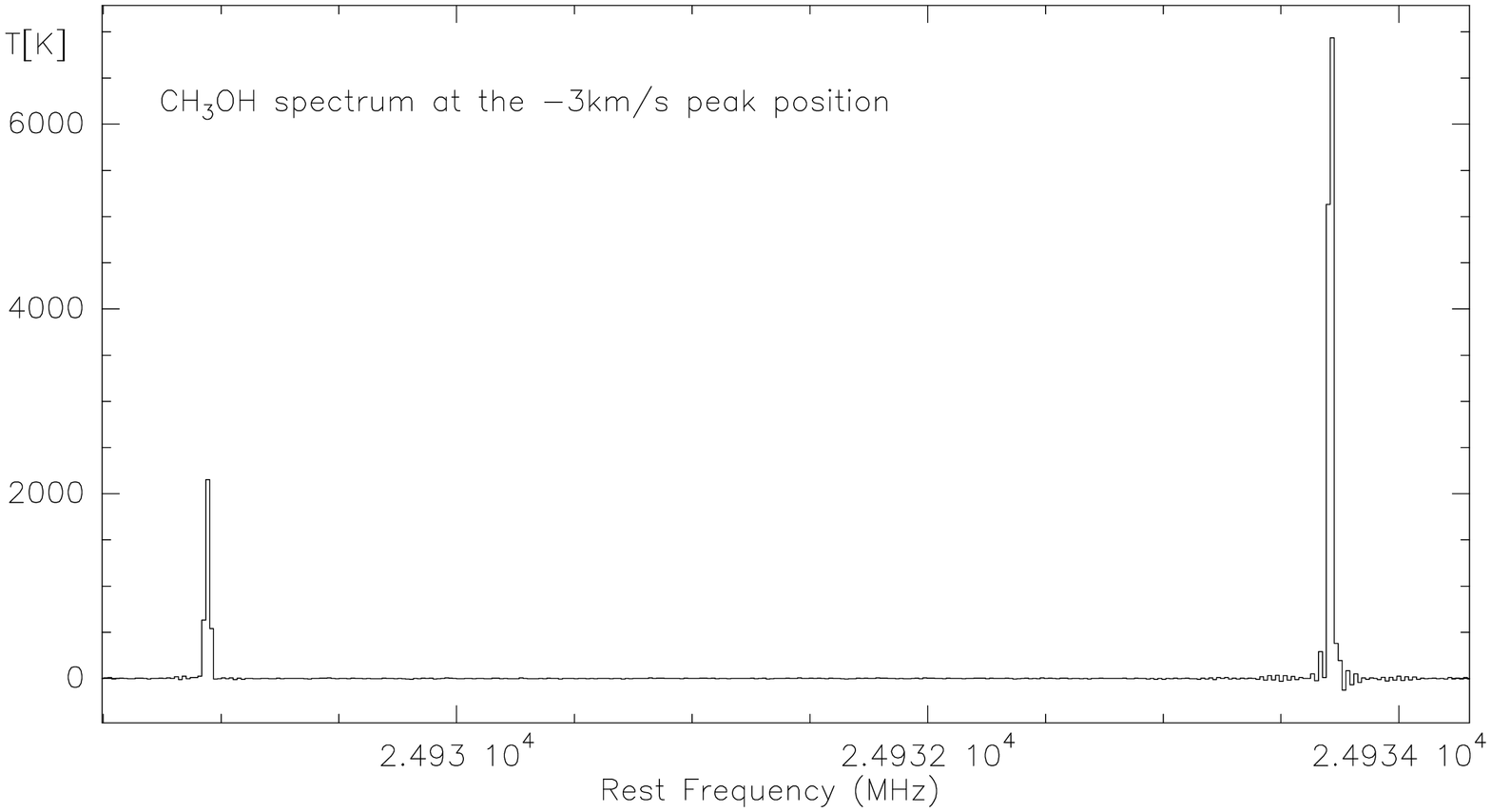}\\
\includegraphics[angle=-90,width=8.5cm]{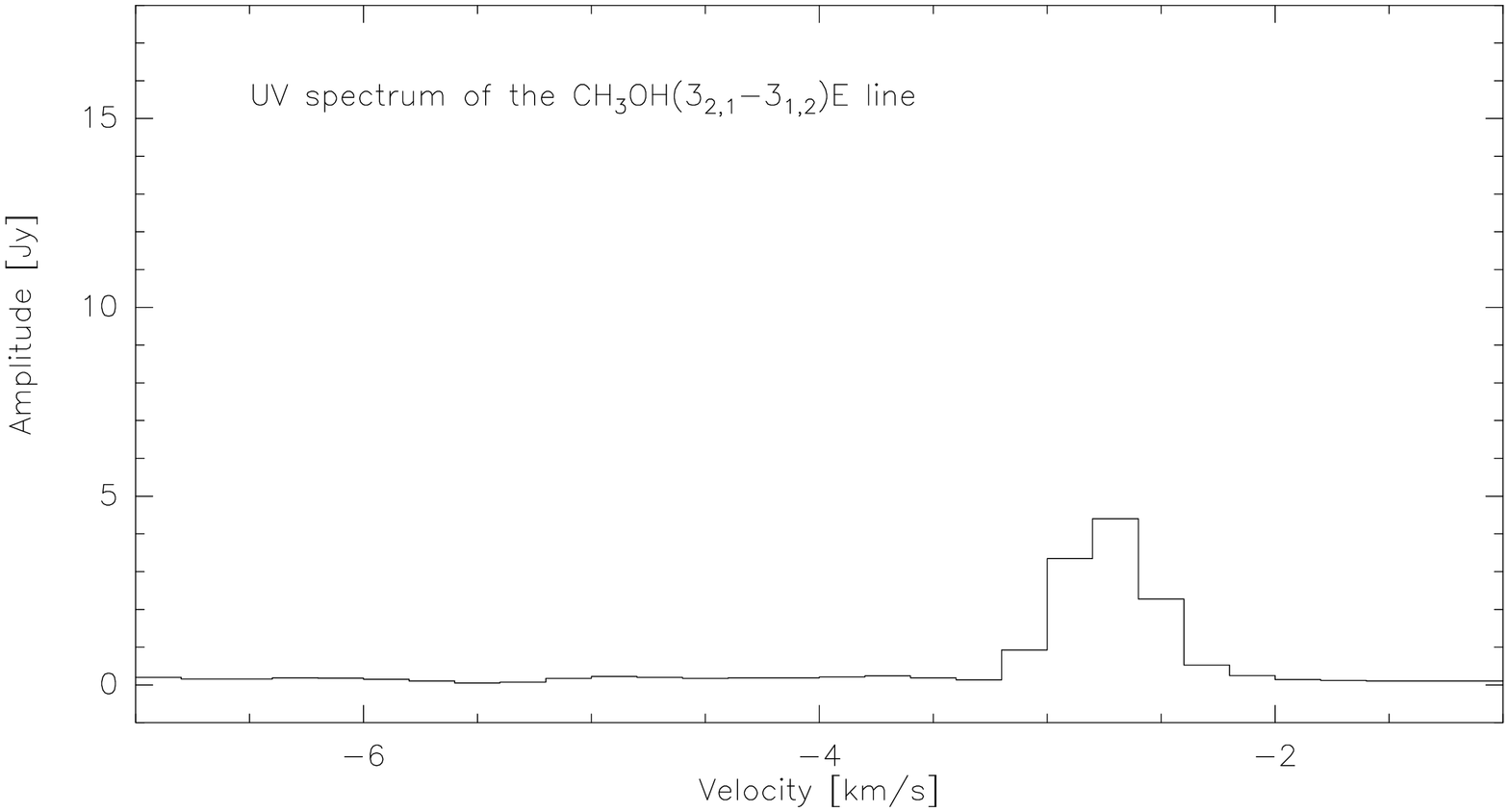}
\includegraphics[angle=-90,width=8.5cm]{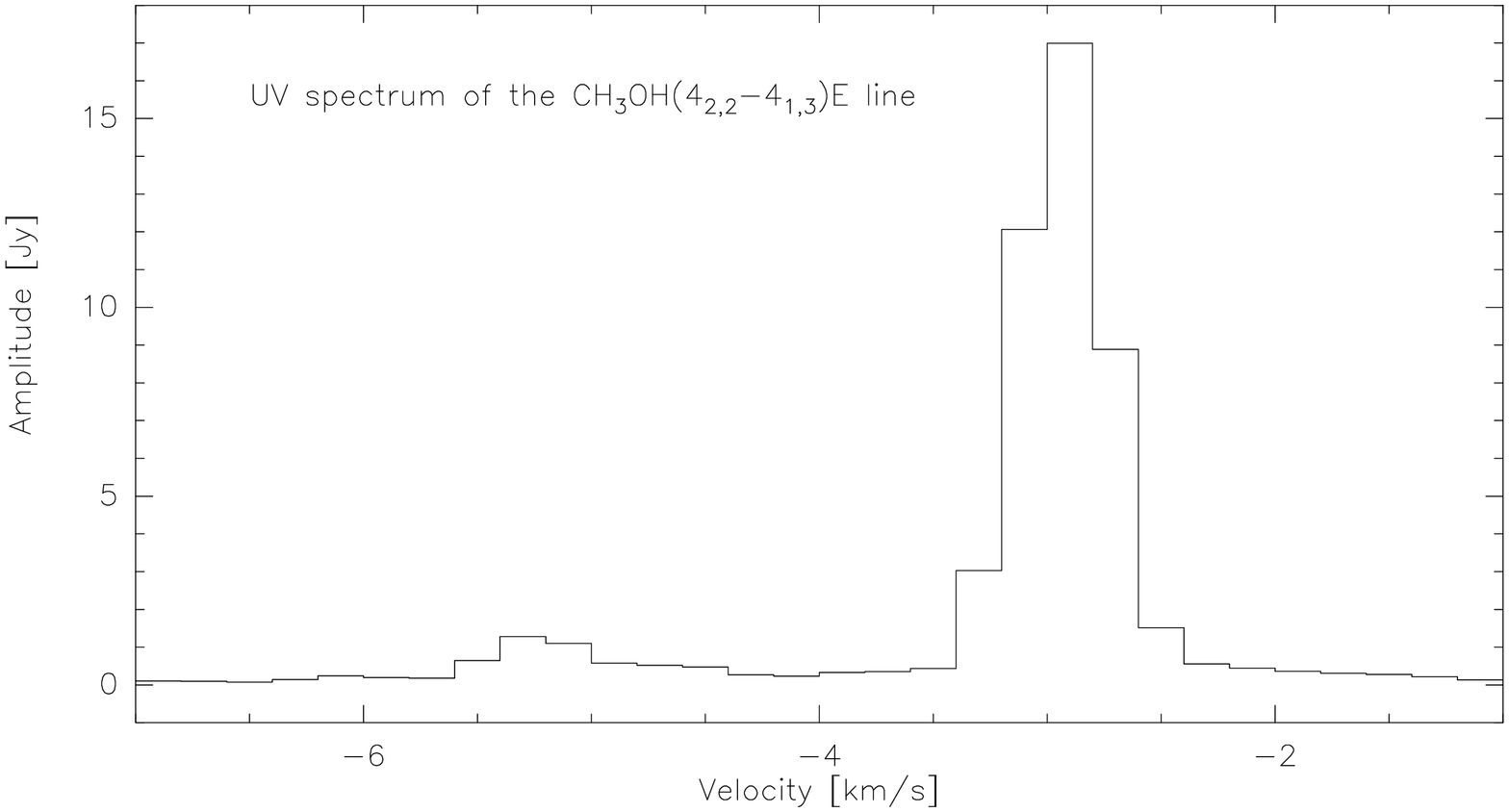}
\caption{CH$_3$OH maser spectra taken toward the NGC6334\,I(N). The
top-left spectrum shows the uv-spectrum on the shortest baseline of
31\,m presenting all detected components. The top-right spectrum is
taken after imaging toward the north-eastern CH$_3$OH maser position
(the conversion factor for K/(Jy beam$^{-1}$) of the right spectrum is
$\sim 385$). The two bottom spectra are the same uv-spectra as the
top-left one but this time presented on the velocity-scale and
differentiated between the \methone~and \methtwo~lines (bottom-left
and bottom-right), respectively.}
\label{ngc6334in_ch3oh_spectrum} \end{figure}

While we do not detect any thermal CH$_3$OH emission, we find Class I
CH$_3$OH maser emission from the \methone~and \methtwo~lines (nothing
from the \meththree~line). Both lines show a maser component at $\sim
-3$\,km\,s$^{-1}$, but the \methtwo~line exhibits another maser
component at $\sim -5$\,km\,s$^{-1}$
(Fig. \ref{ngc6334in_ch3oh_spectrum}). The spatial position of the
$-3$\,km\,s$^{-1}$ components are approximately $7''$ north-east of
the mm-peak in the direction of the collimated SiO outflow, whereas
the $-5$\,km\,s$^{-1}$ component is located about $10''$ south of the
mm-peak (the exact positions are given in Table \ref{maserpos}). The
observed peak brightness temperature of the $-3$\,km\,s$^{-1}$ line is
$\sim 7000$\,K (Fig. \ref{ngc6334in_ch3oh_spectrum}). The
$-5$\,km\,s$^{-1}$ component is less clearly associated with one of
the two outflow axes, but both components are spatially close to
previously detected Class I CH$_3$OH$(7_0-6_1)\,A^+$ 44\,GHz maser
positions \citep{kogan1998}. We note that there are significantly
fewer, just two, 25\,GHz maser components detected than 44\,GHz
components.  The 23 maser spots observed in the latter line are spread
over an area of $23''\times 32''$, which is somewhat larger, but
comparable to the $\sim 15''$ separation of the 25\,GHz
components. The positional offsets between the 44\,GHz and the 25\,GHz
maser positions may be due to the loss of absolute positional
information during self-calibration of the 44\,GHz data by
\citet{kogan1998}. Performing Gaussian fits to the CH$_3$OH data in
the image plane, we get slightly extended features compared to the
beam size. This indicates unresolved spatial blending of different
maser components within each channel width of 0.2\,km\,s$^{-1}$.

\section{Analysis}

\subsection{Temperature estimates from the NH$_3$ observations}
\label{temperatures}

\subsubsection{The NGC6334\,I peak region} 
\label{ngc6334i_peak}

\begin{figure}
\includegraphics[angle=-90,width=8.5cm]{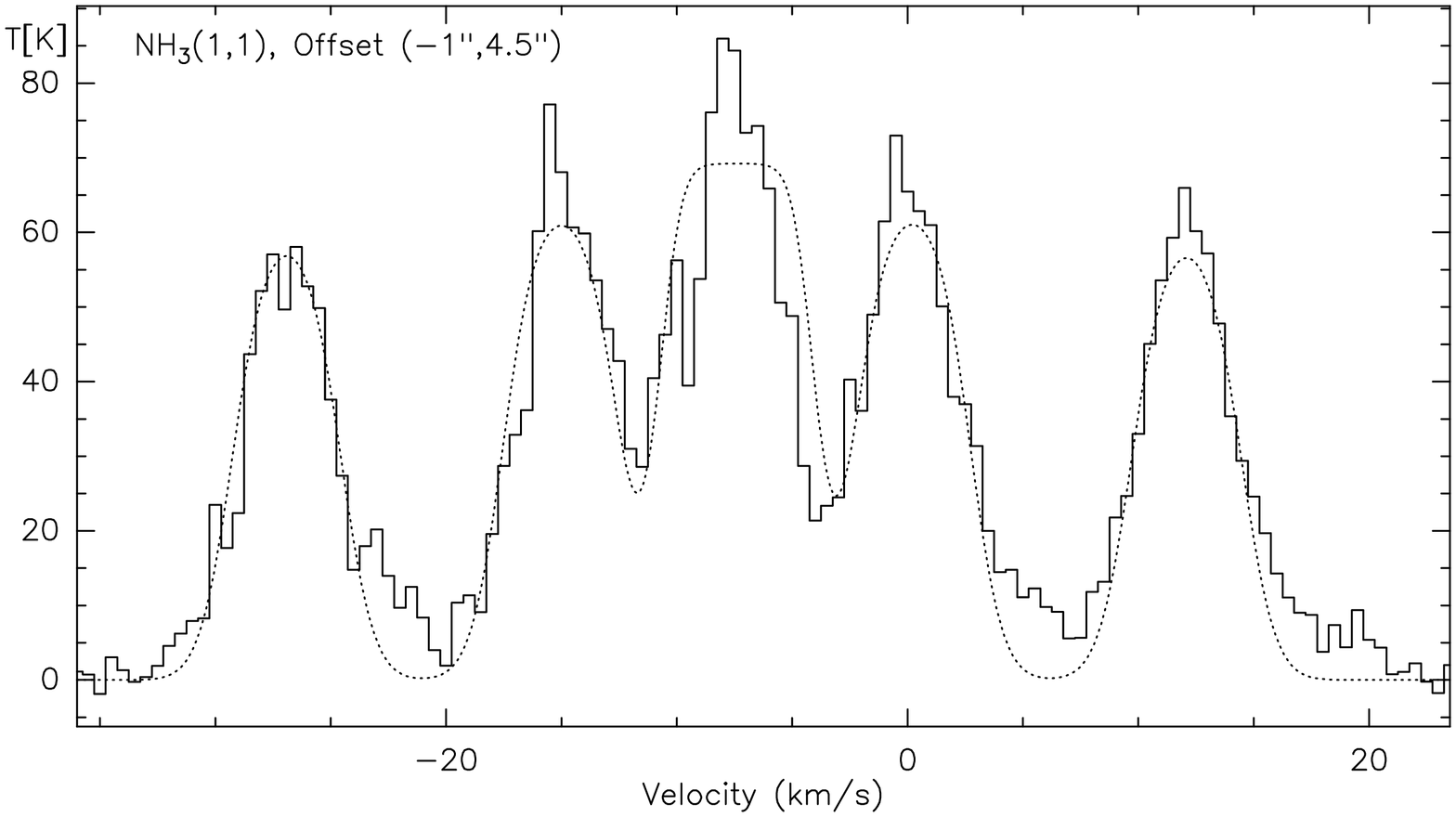}
\includegraphics[angle=-90,width=8.5cm]{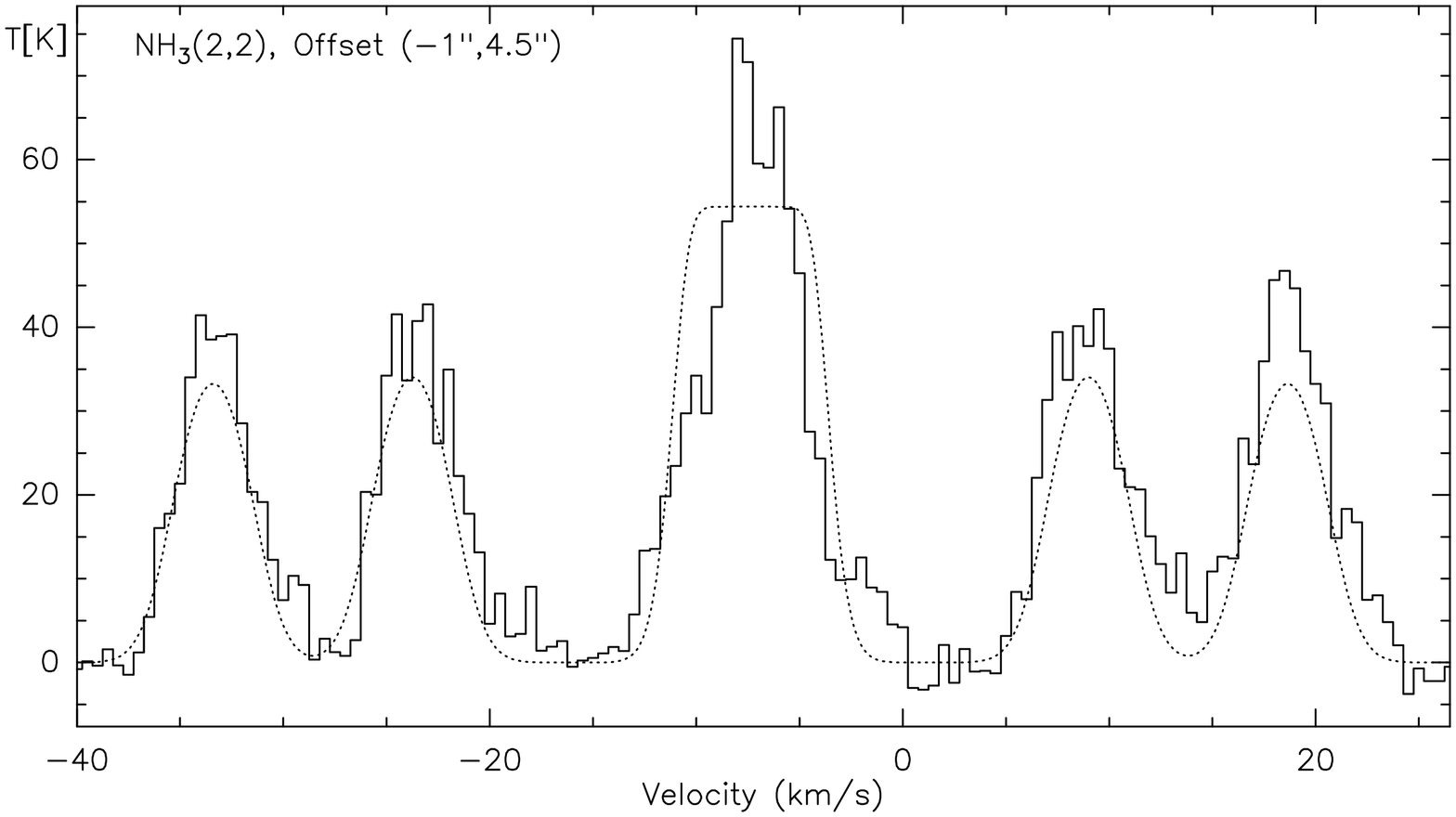}\\
\includegraphics[angle=-90,width=8.5cm]{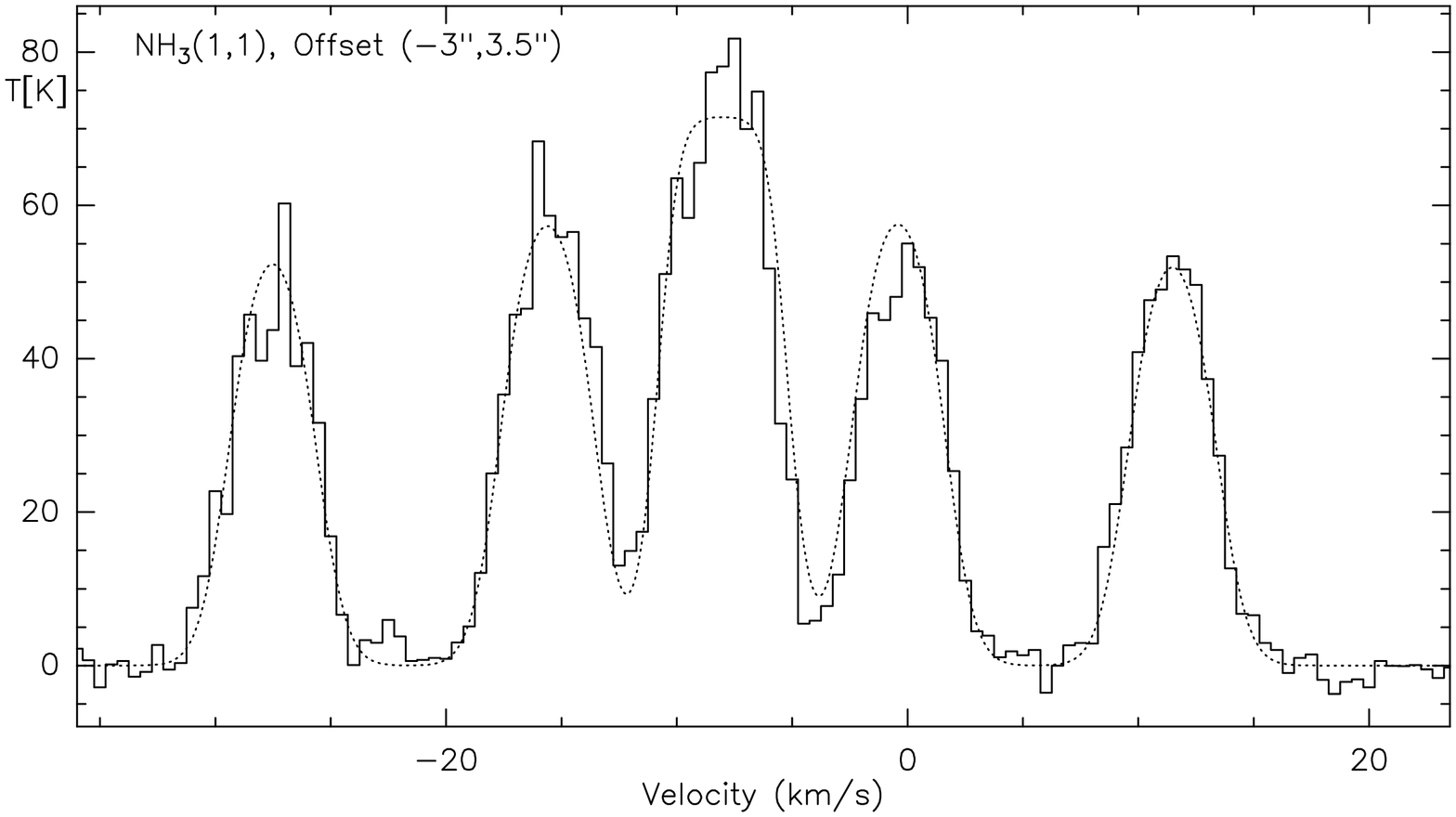}
\includegraphics[angle=-90,width=8.5cm]{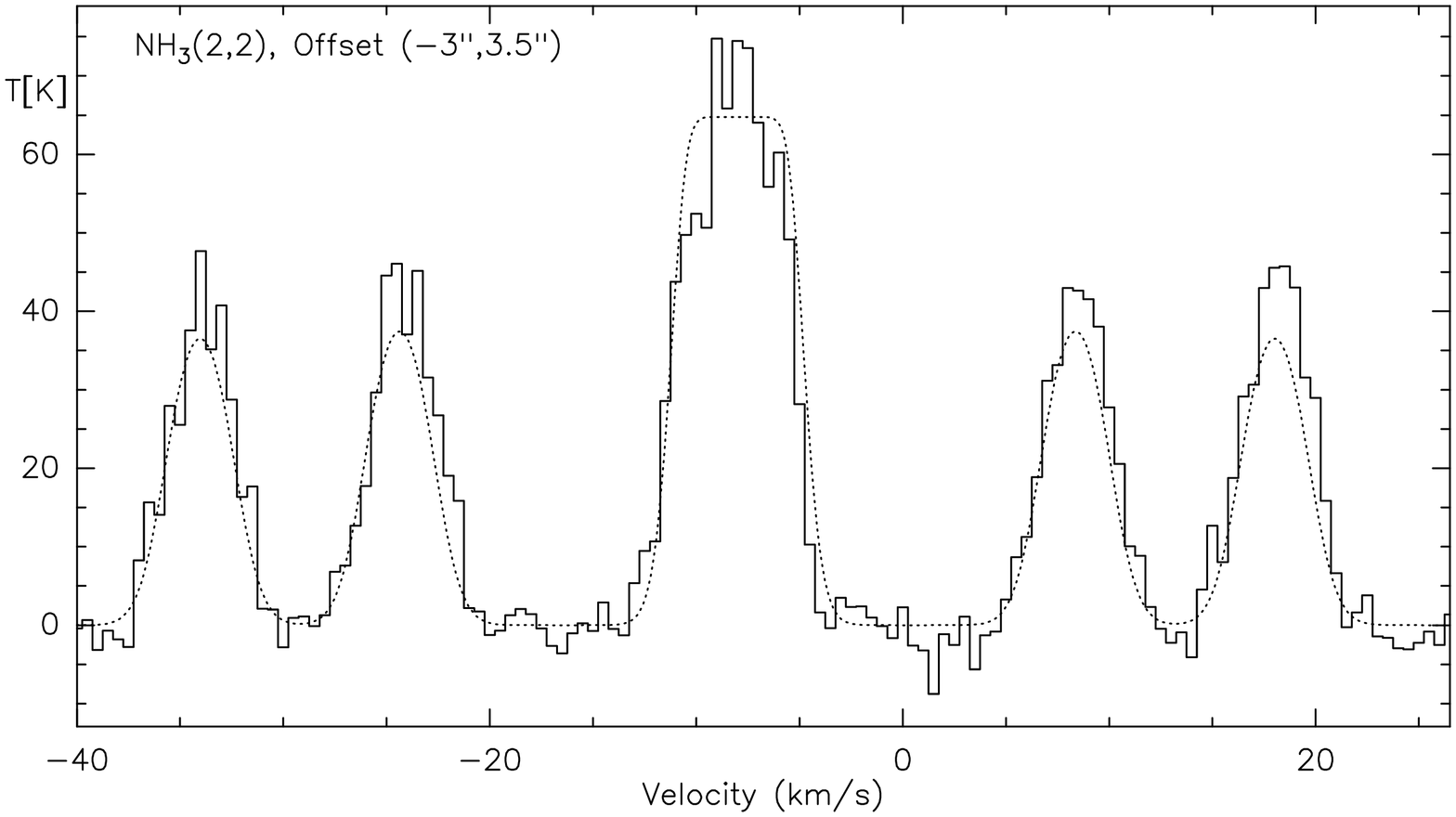}
\caption{NH$_3$ spectra toward the two peak positions in NGC6334\,I
(the offsets of the phase reference centers listed in Table
\ref{center} are marked in each panel). The rather poor fits are shown
in dotted lines.}
\label{ngc6334i_nh3_spectra}
\end{figure}

\begin{figure}
\includegraphics[angle=-90,width=16.6cm]{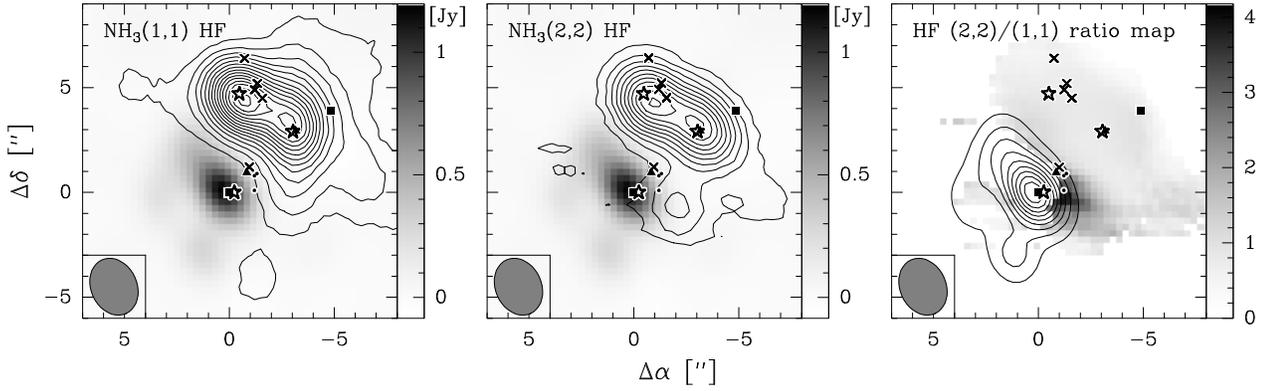}
\caption{Integrated images of the outer satellite hyperfine components
at the most negative velocities toward NGC6334\,I. The left and middle
panels show satellite hyperfine components of NH$_3$(1,1) (the blend
of the NH$_3$(1,1)$F,F_1=0.5,0-0.5,1$ \& $F,F_1=0.5,0-1.5,1$ lines at
$\sim 23.696$\,GHz) and (2,2) (the NH$_3$(2,2)$F=2-1$ line at $\sim
23.725$\,GHz)), respectively. The contour levels for the NH$_3$(1,1)
\& (2,2) satellite components are in $3\sigma$ steps with
$3\sigma (1,1)\sim 10$\,mJy\,beam$^{-1}$ and $3\sigma (2,2)\sim
9$\,mJy\,beam$^{-1}$. The right panel presents a ratio map of the two
integrated satellite hyperfine images (2,2)/(1,1). The cm continuum
emission is presented in grey-scale in the left and middle panels and
in contours (10 to 90\% from the peak emission of 1197\,mJy\,beam$^{-1}$) in
the right panel. The symbols are the same as in Figure
\ref{ngc6334i_images}.}
\label{ngc6334i_hf}
\end{figure}

NH$_3$(1,1) and (2,2) lines are widely used as thermometers in
molecular clouds (e.g., \citealt{walmsley1983,danby1988}). Figure
\ref{ngc6334i_nh3_spectra} shows NH$_3$(1,1) and (2,2) spectra toward
the two NH$_3$ peak positions overlayed with fits to the hyperfine
structure. The hyperfine structure line fits were performed in CLASS
(part of the GILDAS software package) assuming a Gaussian velocity
distribution, equal excitation temperatures and the theoretical
intensity ratio and spacing between main and satellite hyperfine
lines. The fitted parameters are the excitation temperature, the
opacity of the main hyperfine component, the line-width and the
$v_{lsr}$. Obviously, the flat-topped fits do not reproduce the
observed spectra well which is likely due to an extremely high optical
depth and varying excitation conditions over the width of the
lines. Missing flux due to missing short spacings does not play a
significant role since we sample large-scale emission in NGC6334\,I(N)
(Fig.~\ref{ngc6334in_images}) and the uv-coverage of both observations
is nearly the same. These fits do not allow us to derive accurate
temperatures. Therefore, we tried to work only with the satellite
hyperfine components of the spectra, and Figure \ref{ngc6334i_hf}
presents the integrated emission maps of the outer satellite hyperfine
components at the most negative velocities (the blend of the
NH$_3$(1,1)$F,F_1=0.5,0-0.5,1$ \& $F,F_1=0.5,0-1.5,1$ lines at $\sim
23.696$\,GHz and the NH$_3$(2,2)$F=2-1$ line at $\sim
23.725$\,GHz). In the following, we refer to these lines as the
(outermost) satellite hyperfine (hf) components. An intriguing
difference between the integrated NH$_3$(1,1) and (2,2) satellite
hyperfine structure line maps is that the (1,1) satellite hf-line
shows a similar peak intensity distribution as the (1,1) main hf-line
whereas the (2,2) satellite hf-line shows an additional intensity peak
a few arc-seconds south directly at the western edge of the UCH{\sc
ii} region peak. A (2,2)/(1,1) hf-ratio-map between the two integrated
hf-maps stresses this difference as well (Figure \ref{ngc6334i_hf},
right panel). The peak NH$_3$(2,2)/(1,1) hf-ratios at this position
are $\sim 4$ in contrast to ratios of $\sim 0.8$ toward the two main
peak positions.

The energy levels of the NH$_3$(1,1) and (2,2) lines are 23 and
64\,K above ground, and this line-pair is mainly sensitive to
temperatures below 50\,K. As shown by \citet{danby1988}, rotational
temperatures derived from NH$_3$(1,1) and (2,2) line observations are
not sensitive to kinetic temperature changes above about
50\,K. Therefore, NH$_3$(2,2)/(1,1) hf-ratios $> 0.4$ are of no use
for temperature estimates. This is the case for both mm-peak positions
as well as the additional (2,2) hf-peak close to the UCH{\sc ii}
region (Fig.\,\ref{ngc6334i_hf}). However, we observe NH$_3$(1,1) peak
brightness temperatures $> 70$\,K toward the two mm peak positions,
and these brightness temperatures can be considered as lower limits
for the kinetic temperatures at these positions. As the (2,2)/(1,1)
hf-ratio toward the additional (2,2) hf-peak is significantly higher
($\sim 4$), the temperatures there should be higher as well. While
it is likely that massive protostellar objects at the center of the
two mm-peaks are causing the high temperatures there, the hot spot
close to the UCH{\sc ii} region is probably being caused by heating of
the UCH{\sc ii} region itself.

The energy levels above ground for the three CH$_3$OH lines are
between 28 and 44\,K, thus these lines cannot be used for more
accurate temperature determinations as well.

\subsubsection{The NGC6334\,I outflow region} 

The situation is different in the outflow region, and we can extract
reasonable NH$_3$(1,1) and (2,2) spectra toward the NH$_3$(1,1)
outflow-associated peak at offsets ($-10'',-6''$). Fitting the whole
hyperfine spectra at this position, we can derive a rotational
temperature of $T_{\rm{rot}}\sim 19$\,K (for details on the
temperature calculation from NH$_3$ observations, see, e.g.,
\citealt{ungerechts1986}). At these low temperatures, rotational
temperatures correspond well to kinetic temperatures
\citep{danby1988}.

\subsubsection{The northern source NGC6334\,I(N)} 

\begin{figure}
\includegraphics[angle=-90,width=8.5cm]{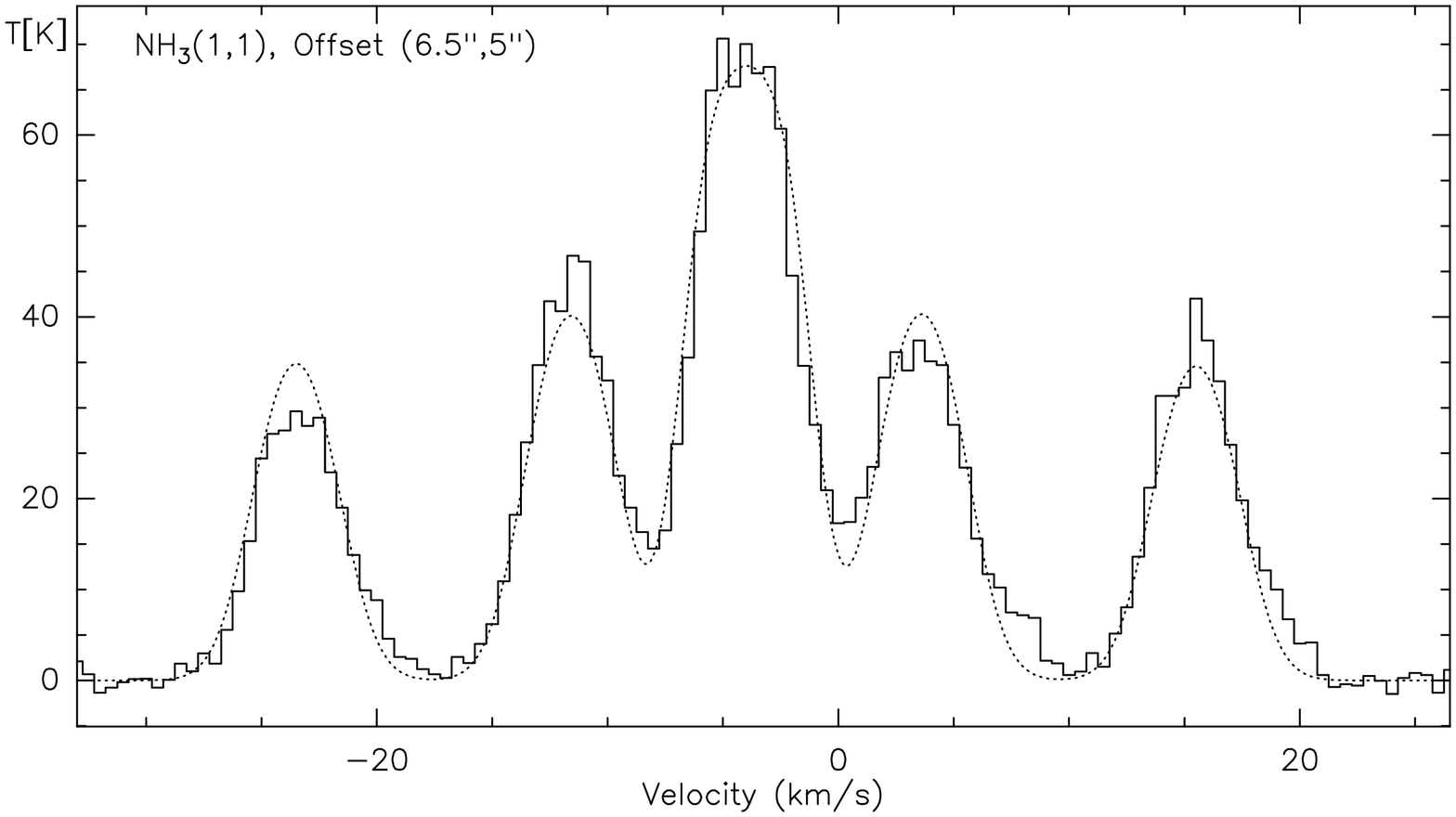}
\includegraphics[angle=-90,width=8.5cm]{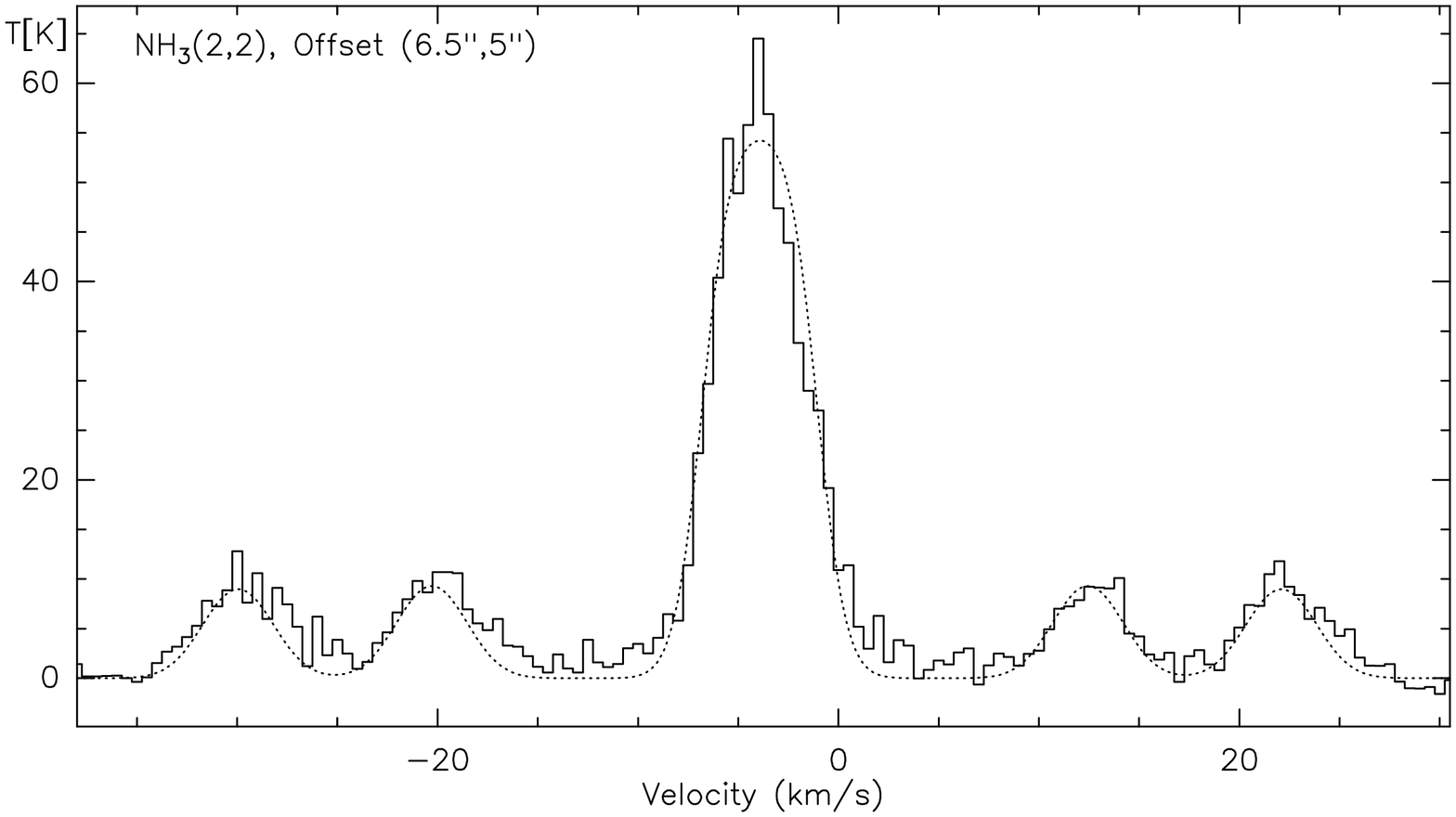}
\caption{NH$_3$(1,1) and (2,2) spectra taken toward the nominal
position of the mm continuum peak discovered by Hunter et al.~(in
prep.) toward NGC6334\,I(N). The offsets of the phase reference
centers listed in Table \ref{center} are marked in each panel. The
fits are shown in dotted lines.}
\label{ngc6334in_nh3_spectra}
\end{figure}

Since we have no well defined NH$_3$ peak position in NGC6334\,I(N), we
extracted spectra toward the mm peak position determined by Hunter et
al.~(in prep.), which is offset by $(6.5'',5'')$ from our phase
reference center. The brightness temperatures are comparable to the
brightness temperatures in the southern source NGC6334\,I but the
opacities are lower as observed in the lower hyperfine line strengths
(especially in the (2,2) lines, Figure
\ref{ngc6334in_nh3_spectra}). Therefore, we can fit these lines
reasonably well and again derive rotational temperatures as done
previously for the outflow position. The derived NH$_3$ rotational
temperature is $T_{\rm{rot}}\sim 47$\,K corresponding to kinetic
temperatures $T_{\rm{kin}}>100$\,K \citep{danby1988}. Thus, again we
can derive only lower limits for the temperatures toward
NGC6334\,I(N). This estimate based on the high-spatial resolution
interferometric observations is significantly higher than previous
estimates of 30\,K from single-dish NH$_3$ observations with a beam
size of $55''$ \citep{kuiper1995}, because the latter observations
averaged over the whole region.

\subsection{Kinematic structures of the molecular gas in NGC6334\,I}

\begin{figure}
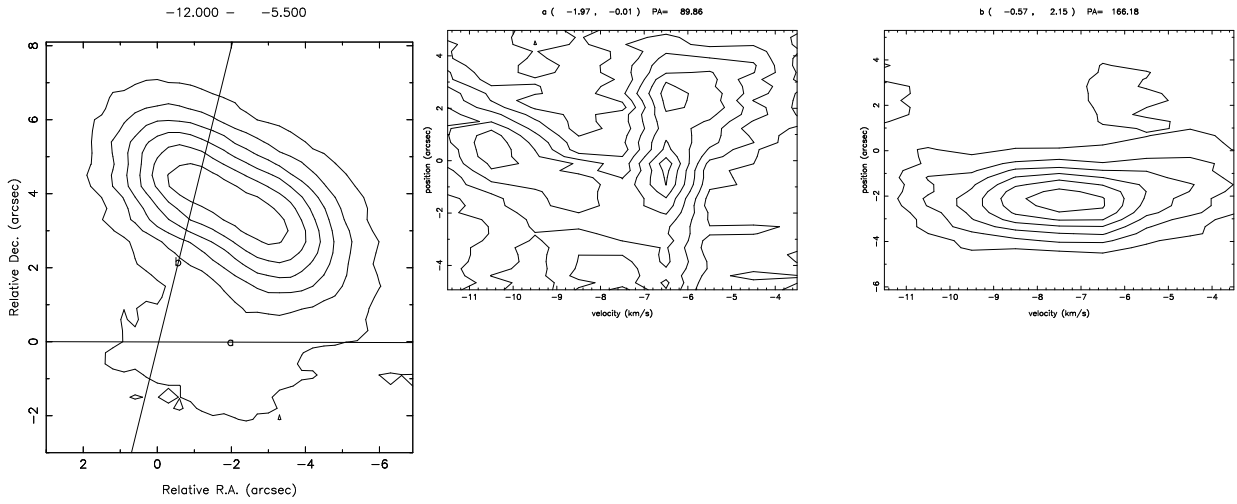

\includegraphics[angle=-90,width=5.4cm]{f7a.ps}
\includegraphics[angle=-90,width=10.8cm]{f7b.ps}
\caption{Position-velocity cuts through the NH$_3$(2,2) outer
satellite hyperfine line data cube (NH$_3$(2,2)$F=2-1$, integrated
from $-12$ to $-5$\,km\,s$^{-1}$) of NGC6334\,I along the axes
outlined in the left panel. The middle panel corresponds to the cut in
east-west direction, the right panel shows the cut north-west
south-east direction. The contour levels are in 11.5, 5.0 and
16.5\,mJy\,beam$^{-1}$ steps for the left, middle and right panel,
respectively.}
\label{ngc6334i_hf_pv}
\end{figure}

The striking edge-on orientation between the UCH{\sc ii} region and
the molecular peaks in NGC6334\,I allows a closer analysis of possible
kinematic signatures the UCH{\sc ii} region might impose on the
molecular cores (in addition to the heating effect we have described
in \S \ref{ngc6334i_peak}). Figure \ref{ngc6334i_hf_pv} shows two
position-velocity (pv) cuts performed for the NH$_3$(2,2): (a) in
east-west direction from the UCH{\sc ii} region through the
NH$_3$(2,2) hyperfine emission hot spot, and (b) from the UCH{\sc
ii} region in north-western direction through the main mm and NH$_3$
peak position. While there is considerable structure in the east-west
pv-diagram (a), it however does not reveal any clear signature that
the UCH{\sc ii} could have imposed on the molecular gas. The other
pv-diagram (b) shows no strong velocity gradients from the UCH{\sc ii}
region in the direction of the NH$_3$ peak position as well.

\begin{figure}
\includegraphics[angle=-90,width=16.2cm]{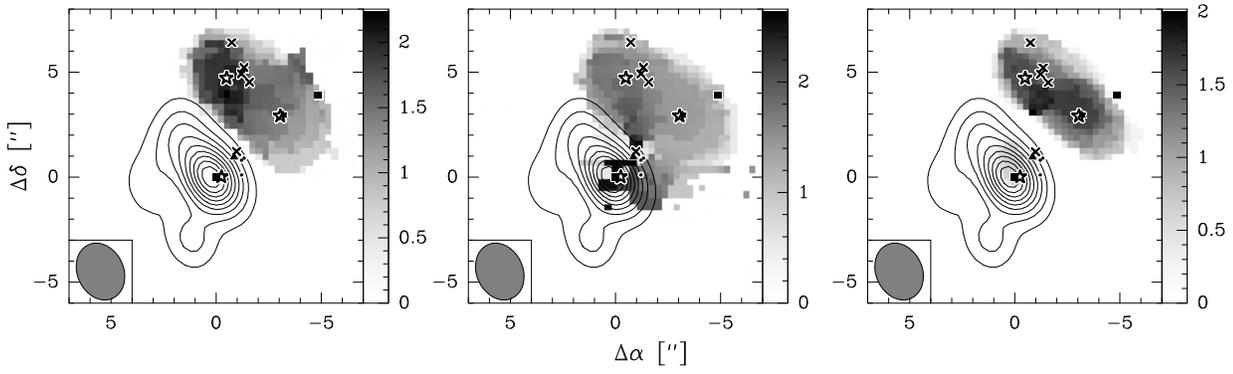}
\caption{The grey-scale shows moment 2 maps of the NH$_3$(1,1) (left),
(2,2) (middle) outermost satellite hyperfine lines and the
CH$_3$OH($3_{2,1}-3_{1,2}$)E line (right panel) in NGC6334\,I (the
wedge is in units of km\,s$^{-1}$). The contours in all three panels
show the cm continuum data from 10 to 90\% of the peak emission of
1197\,mJy\,beam$^{-1}$. The symbols are the same as shown in
Fig.~\ref{ngc6334i_images}.}
\label{ngc6334i_mom2}
\end{figure}

Figure \ref{ngc6334i_mom2} presents the moment 2 maps (velocity
dispersion maps) of the two NH$_3$(1,1) \& (2,2) outermost satellite
hyperfine lines and one of the CH$_3$OH lines. While the velocity
dispersion of the (1,1) hf-line peaks toward the north-eastern NH$_3$
peak, the (2,2) velocity dispersion peaks further south toward the
UCH{\sc ii} region and the NH$_3$(2,2) hyperfine hot spot.  In
contrast, the moment 2 map of CH$_3$OH shows two peaks both associated
with the two molecular peak positions. 

At estimated temperatures of $\sim 100$\,K the critical densities of
the CH$_3$OH lines are of the order $10^4$\,cm$^{-3}$, whereas the
critical densities of the NH$_3$ lines are lower (of order
$10^3$\,cm$^{-3}$). The velocity dispersion of the higher density
tracer CH$_3$OH is dominated by internal motion due to the embedded
protostars. To some degree this is also the case for NH$_3$, and
especially the (1,1) moment 2 maps peaks toward the strongest mm
continuum peak. However, there is no obvious (1,1) velocity dispersion
peak toward the secondary mm peak position. The NH$_3$(2,2) velocity
dispersion is clearly dominated by the NH$_3$(2,2) hyperfine hot spot
close to the UCH{\sc ii} region.  Since NH$_3$ has lower critical
densities than CH$_3$OH, its line emission shows comparably less
contribution from the densest cores, and it is thus more likely to
exhibit temperature and kinematic signatures imposed by the adjacent
UCH{\sc ii} region.

\subsection{Molecular absorption toward the UCH{\sc ii} region}

\begin{figure}
\includegraphics[angle=-90,width=5.4cm]{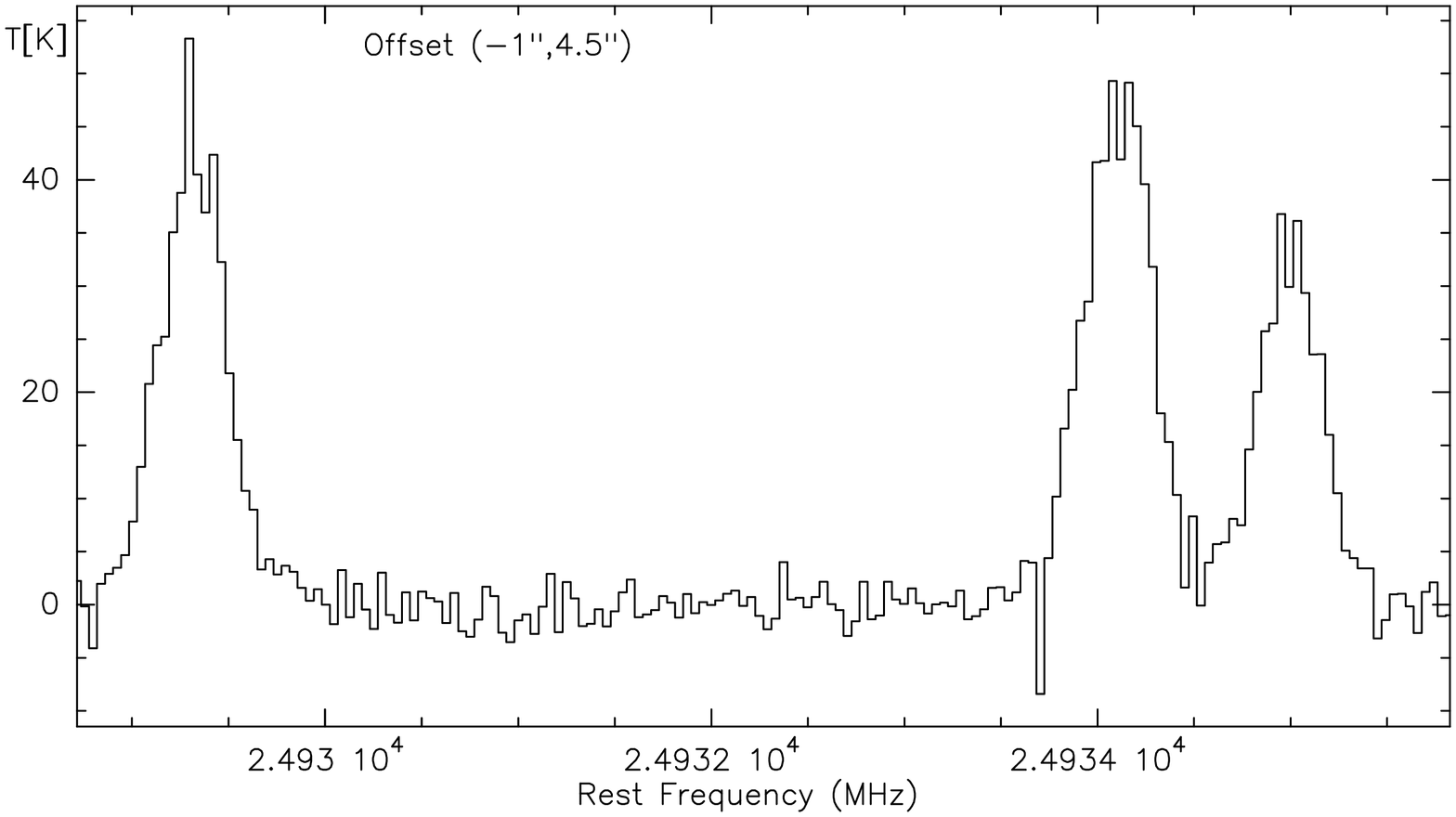}
\includegraphics[angle=-90,width=5.4cm]{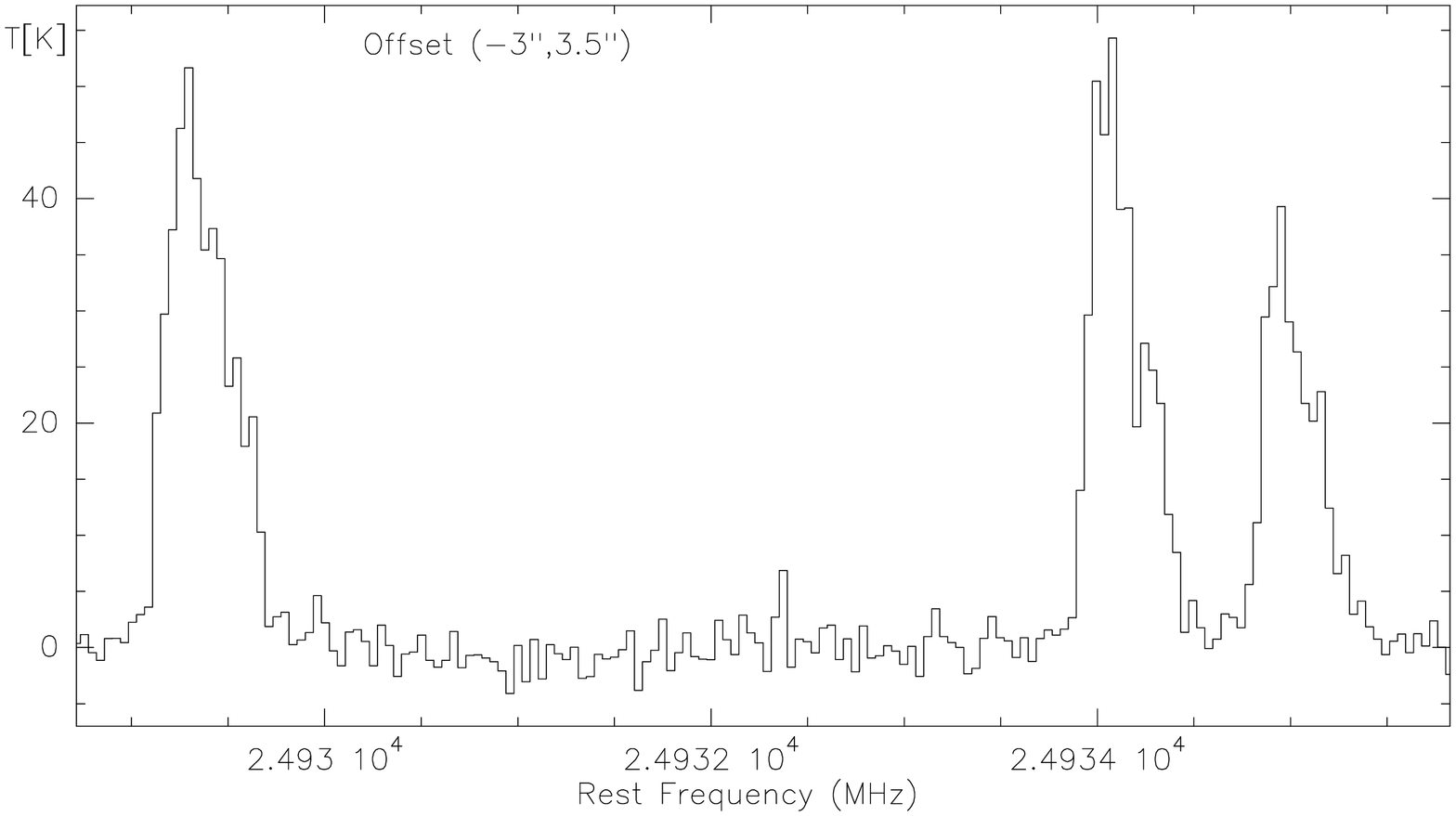}
\includegraphics[angle=-90,width=5.4cm]{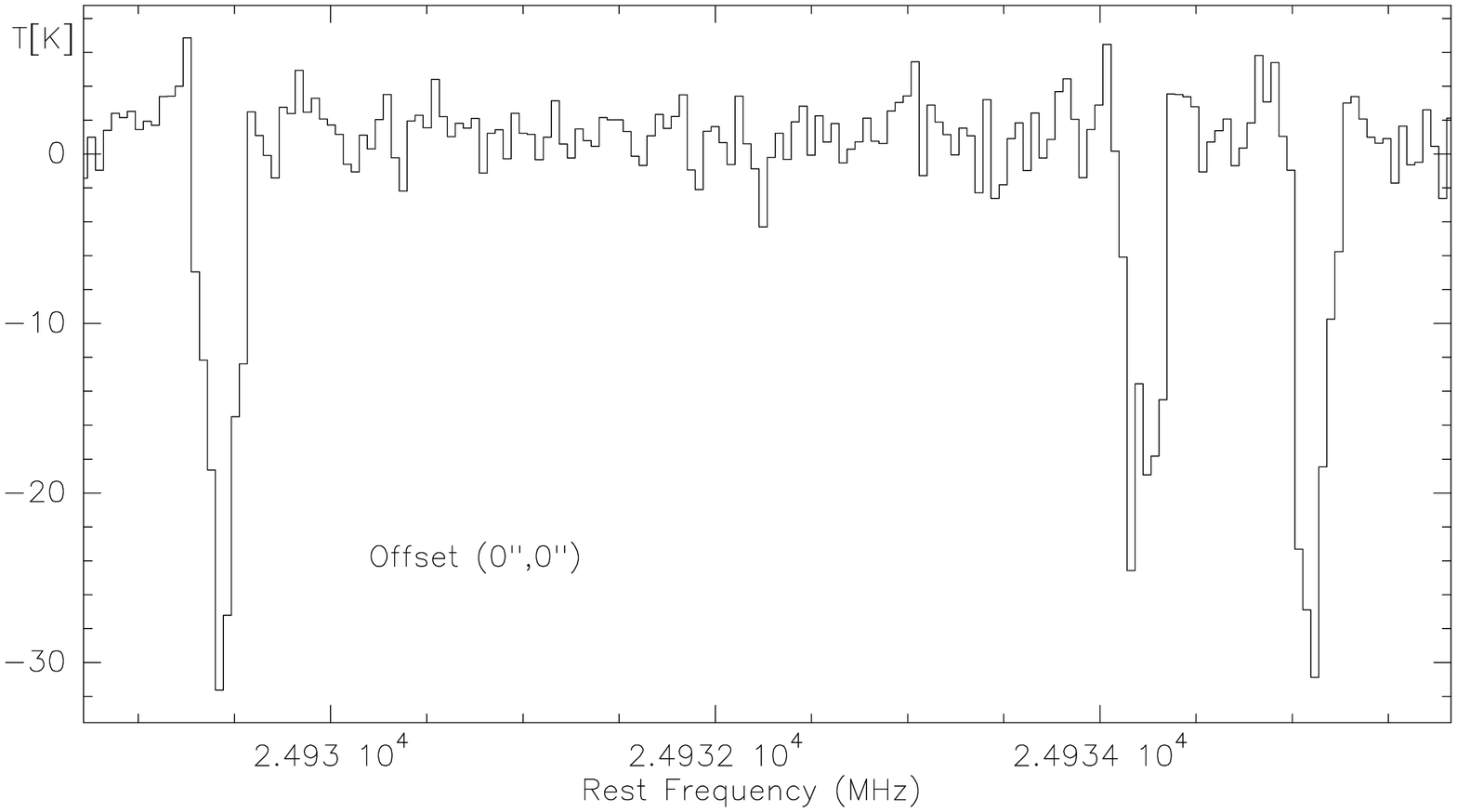}
\caption{CH$_3$OH spectra taken toward the two emission peaks and the
UCH{\sc ii} region (right panel) in NGC6334\,I. The offsets are given
with respect to the phase center listed in Table \ref{center}.}
\label{ngc6334i_ch3oh_spectra}
\end{figure}

In addition to the molecular line emission, there are prominent
CH$_3$OH line absorption features observed in all three lines toward
the UCH{\sc ii} region in NGC6334\,I
(Fig. \ref{ngc6334i_ch3oh_spectra}). Examining the NH$_3$ data, we
find no absorption in the satellite hyperfine lines but weak NH$_3$
absorption features in the two main lines. Position-velocity cuts of
one of the CH$_3$OH and the two NH$_3$(1,1) and (2,2) main lines
through the UCH{\sc ii} region at $\Delta$Dec.=0 reveal that the
absorption features are blue-shifted with regard to the (weak) emission
features toward the UCH{\sc ii} region. This signature is more
pronounced in the CH$_3$OH and NH$_3$(1,1) line whereas it is not
obviously discernible in the NH$_3$(2,2) line. Nevertheless, these
data indicate expanding molecular gas around the UCH{\sc ii} region.

\section{Discussion}

The differences between the two massive star-forming regions
NGC6334\,I \& I(N) in their molecular NH$_3$ and CH$_3$OH emission are
rather astonishing: compact thermal emission is observed in both
species toward the UCH{\sc ii} region/hot core NGC6334\,I whereas
extended NH$_3$ and masing CH$_3$OH emission is found toward
NGC6334\,I(N). It is not entirely obvious what is causing these
differences because both sources harbor compact mm continuum sources
as well as jet-like SiO outflows (Hunter et al. in prep.). Since the
southern source NGC6334\,I has a UCH{\sc ii} region, it is likely that
evolutionary processes are at least partly responsible for the
different molecular signatures. However, with the data obtained during
the present investigation we still cannot accurately determine
temperature and density in both regions, and one expects that the
molecular emission is sensitive to these parameters. Therefore, it is
an important next step to get thorough estimates of the characteristic
temperatures and densities of both regions to get a deeper
understanding of the physical processes causing the observational
differences.

\subsection{Four stages of massive star formation within 1\,pc}

\begin{figure}
\includegraphics[angle=-90,width=8cm]{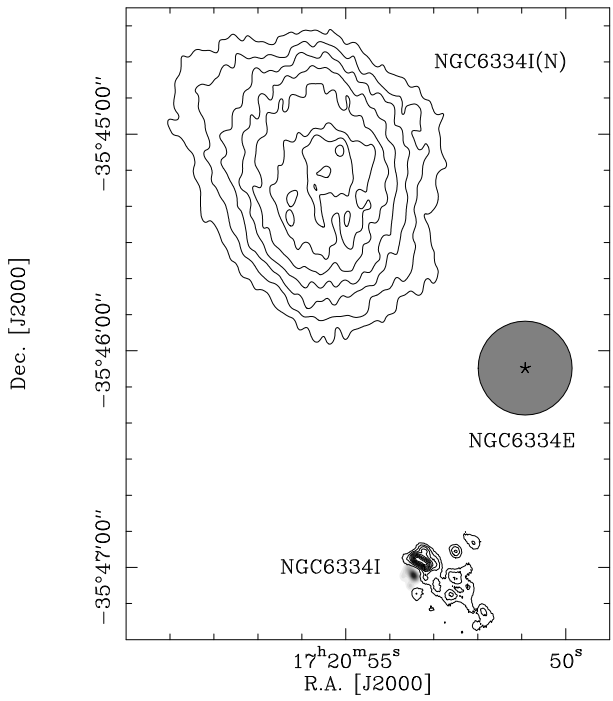}
\caption{Large-scale overview of the northern end of the NGC6334
complex. The grey-scale shows the cm continuum emission of NGC6334\,I
and the contours present the integrated NH$_3$(1,1) emission of the
main hyperfine component. The contour levels are from 10 to 90\% of
the peak values of each image with peak values: cm emission
1197\,mJy\,beam$^{-1}$, NH$_3$(1,1) in NGC6334\,I 175\,mJy\,beam$^{-1}$, NH$_3$(1,1) in
NGC6334\,I(N) 386\,mJy\,beam$^{-1}$. The grey circle and star present the
approximate structure of NGC6334E (e.g., \citealt{carral2002}).}
\label{overview}
\end{figure}

Figure \ref{overview} presents a large-scale overview of the
north-eastern end of the NGC6334 region comprising NGC6334\,I, NGC6334E
and NGC6334\,I(N). It is interesting that within 1\,pc distance we find
four different evolutionary stages of massive star formation: the
shell-like H{\sc ii} region NGC6334E, the champagne-flow-like UCH{\sc
ii} region NGC6334\,I and its adjacent outflow driving mm cores, and
further north the enigmatic source NGC6334\,I(N). NGC6334E lies just
outside each of our primary beams, thus we have no new data about this
source.

From an evolutionary point of view, the shell-like H{\sc ii} region
NGC6334E appears to be the most evolved and thus oldest source within
the region (e.g., \citealt{rodriguez1982,tapia1996,carral2002}). The
UCH{\sc ii} region NGC6334\,I is significantly smaller than NGC6334E and
thus apparently in an earlier evolutionary stage (e.g.,
\citealt{persi1982,depree1995}). Based on the cold dust temperatures,
the detection of only a few very red infrared sources and weak cm
emission \citep{gezari1982,tapia1996,sandell2000,carral2002},
NGC6334\,I(N) was always considered to be the youngest region within the
large-scale NGC6334 complex. However, comparing the mm
continuum/NH$_3$ sources in NGC6334\,I with those in NGC6334\,I(N), it is
not entirely obvious which of these two regions is evolutionary
younger. Both complexes have compact mm dust continuum sources and
drive collimated outflows. In contrast to these similarities, the
differences in the compact versus extended NH$_3$ emission and thermal
versus maser CH$_3$OH emission at 24.9\,GHz are striking. Furthermore,
the differences in the estimated dust temperatures are large with
100\,K for NGC6334\,I and only 30\,K for NGC6334\,I(N)
\citep{sandell2000}. Contrary to this, even constraining only lower
limits for the gas temperatures based on our NH$_3$ observations, we
derive kinetic gas temperatures $\geq 100$\,K for both regions. Even
considering the different beam sizes of the SCUBA observations of the
dust emission (between $8.5''$ and $18.5''$ depending on the
frequency) and our NH$_3$ gas observations, the dust and gas
temperatures in NGC6334\,I appear to be consistent with each other,
whereas this is not the case for NGC6334\,I(N). There, the gas and dust
temperatures appear to have decoupled. Additionally,
\citet{sandell2000} estimated the bolometric luminosities of both
sources to differ by about two orders of magnitude with $2.6\times
10^5$\,L$_{\odot}$ and $1.9\times 10^3$\,L$_{\odot}$ for NGC6334\,I \&
I(N), respectively. This difference could either imply that
NGC6334\,I(N) may not form a cluster as massive as NGC6334\,I, or
NGC6334\,I(N) may be less evolved and younger than NGC6334\,I and thus may
still build up its final larger luminosity. With the observations so
far, we cannot clearly judge which of these two regions is in an earlier
evolutionary stage. With the UCH{\sc ii} region and the infrared
cluster present, the whole region NGC6334\,I is older than the region
NGC6334\,I(N), but it is less clear whether the same conclusion holds
for the mm/NH$_3$ cores within NGC6334\,I.  However, it is worth
mentioning that the non-detection of compact NH$_3$ emission toward
the mm sources in NGC6334\,I(N) is a rarely observed phenomenon and may
indicate a short-lived period during the evolution of massive
star-forming regions.

\subsection{Turbulence versus organized motion}

\begin{figure}
\includegraphics[angle=-90,width=8cm]{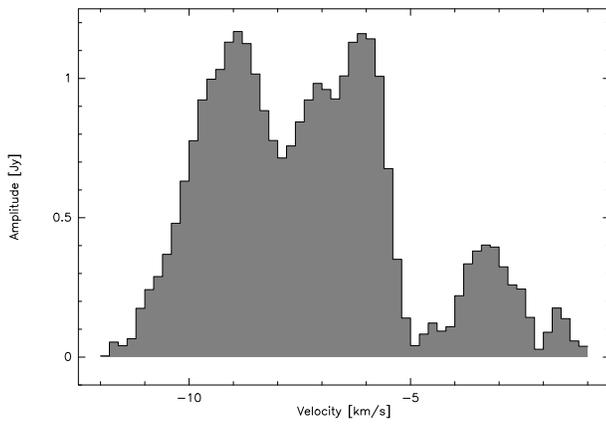}
\caption{NH$_3$(1,1) spectrum of the outer satellite hyperfine
component taken in the uv-domain on the shortest baseline of 31\,m.}
\label{6334i_hf_spectrum}
\end{figure}

\begin{figure}
\includegraphics[angle=-90,width=16cm]{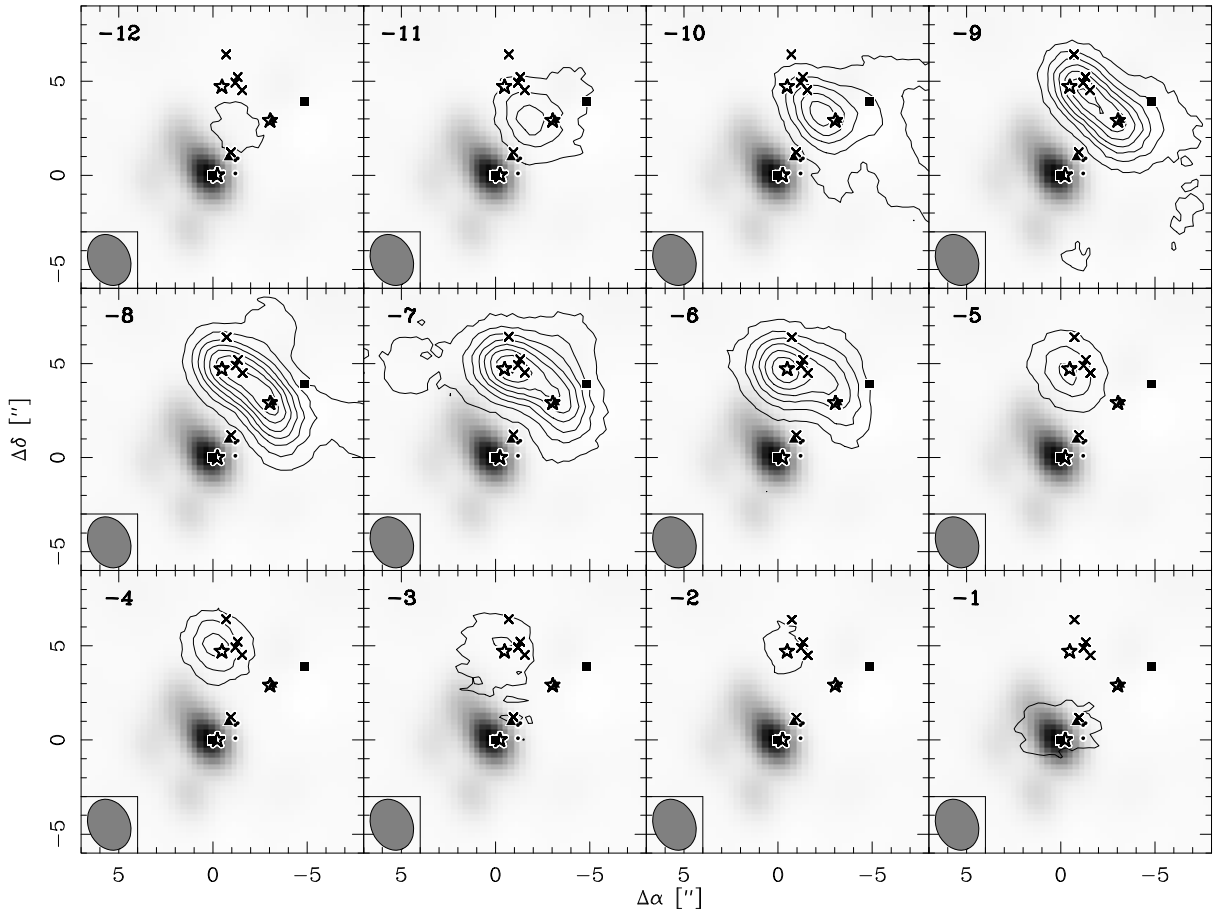}
\caption{The contours present a channel map with 1\,km\,s$^{-1}$
resolution of the NH$_3$(1,1) outer satellite hyperfine component
shown in Figure \ref{6334i_hf_spectrum} (contour levels
(20(160)20\,mJy\,beam$^{-1}$). The central velocities are shown at the
top-left of each panel and the symbols are the same as in
Fig.~\ref{ngc6334i_images}. The grey-scale always show the cm
continuum emission.}
\label{6334i_hf_channel}
\end{figure}

Figure \ref{6334i_hf_spectrum} shows a spectrum of the outermost
satellite hyperfine component of the NH$_3$(1,1) line in the uv-domain
on the shortest baseline of 31\,m. Several velocity components are
present: one centered at $\sim -9$\,km\,s$^{-1}$, one at $\sim -6$\,km\,s$^{-1}$, one
at $\sim -3$\,km\,s$^{-1}$, and one at $\sim -1.5$\,km\,s$^{-1}$. Analyzing the
spatial distribution of the different velocity components (Figure
\ref{6334i_hf_channel}) we find that only the $-9$\,km\,s$^{-1}$ component is
clearly associated with the south-western NH$_3$ emission peak,
whereas all other features are associated with the north-eastern peak
(which is also the peak of stronger mm continuum emission).

Contrary to the complex profile in the uv-domain (Figure
\ref{6334i_hf_spectrum}), the hyperfine components of the spectra
taken individually toward each peak position resemble single-peaked
Gaussians (Figures \ref{ngc6334i_nh3_spectra} \&
\ref{ngc6334in_nh3_spectra}). We conducted Gaussian line-fits to the
outermost satellite hyperfine components of the NH$_3$(1,1) line
toward the two NH$_3$ peak position in NGC6334\,I and the mm peak
position in NGC6334\,I(N), and the derived line-widths $\Delta v$ vary
between 4.1 and 5.1\,km\,s$^{-1}$ (see Table \ref{linewidths}).

What is causing these broad line-widths, turbulent motion due to,
e.g., molecular outflows, or organized motion due to, e.g., infall,
envelope rotation, disks, or multiple components? \citet{sridha} have
shown that the NH$_3$ linewidth observed with single-dish telescopes
toward High-Mass Protostellar Objects (HMPOs) is significantly lower
than toward typical UCH{\sc ii} regions and hot cores ($\Delta
v_{\rm{HMPO}} \sim 2.1$\,km\,s$^{-1}$ versus $\Delta v_{\rm{hot core}}
\sim 3.1$\,km\,s$^{-1}$). Since we find gas temperatures of the order
100\,K, NGC6334\,I \& I(N) are considered to be part of the hot core
group.  Therefore, most likely the large line-widths are an
evolutionary signature of the hot core stage and no remnant from the
original molecular core. The thermal linewidth at 100\,K is only
0.31\,km\,s$^{-1}$ and thus non-thermal motion has to account for
nearly all the observed linewidth. To estimate the corresponding
energy within the core we calculated $E=\frac{1}{2}m\,\Delta
v_{\rm{nt}}^2$ with $m$ the core mass derived from the mm continuum
emission (Hunter et al.~in prep.)  and $\Delta v_{\rm{nt}}
=\sqrt{\Delta v^2 - \Delta v^2_{\rm{thermal}}}$ the non-thermal
linewidth component. The derived energy values corresponding to
internal motion are of the order $10^{46}$\,erg (see Table
\ref{linewidths}).

Interestingly, the typical energy supply of massive molecular outflows
is of the same order (e.g., \citealt{beuther2002b}). The spatial
scales for the molecular outflows are usually of the order 1 parsec
whereas the NH$_3$ line-widths observed toward NGC6334\,I are confined
to a few thousand AU, thus the estimated energies are not directly
comparable. However, the uncertainties in deriving outflow energies
are usually high, up to an order of magnitude (e.g.,
\citealt{cabrit1990}), and it might be possible that comparable
amounts of energy could be supplied from the outflow on spatial scales
as small as the NH$_3$ cores. Therefore, turbulence caused by
molecular outflows may contribute to the observed linewidth in
NGC6334\,I. Contrary to this, \citet{zhang1998a,zhang2002} find broad
NH$_3$ line-widths of the order 5\,km\,s$^{-1}$ toward two young massive
star-forming regions, whereas the NH$_3$ emission in the surrounding
cores has a narrower linewidth of the order 2\,km\,s$^{-1}$. These
observations are interpreted as indirect evidence for rotating disks
within these massive star-forming regions. Since we are unable to
derive any systematic velocity gradient within the separate cores of
NGC6334\,I, we cannot attribute our observations to a disk
scenario. However, it is likely that the observed line-widths are
caused at least to some degree by organized motion within the inner
few thousand AU caused by either infall, envelope rotation, disks or
multiple components, or an interplay of various of these
components. Only higher angular resolution observations can
disentangle all these contributions.

\subsection{Shocks in NGC6334\,I(N)}

In contrast to NGC6334\,I, the situation appears to be different in
NGC6334\,I(N) because the NH$_3$ emission is not confined to the mm peak
position, and we observe broad line-widths over a large spatial
area. Since there are at least two molecular outflows in the region,
the broad linewidth could well be caused by shock interactions of the
outflow with the surrounding molecular core. 

Additional evidence for outflow-shock interactions is the detection
and the spatial spread of the Class I 25 and 44\,GHz CH$_3$OH maser
emission which is believed to be caused by the interactions of
outflows from massive star-forming regions with the surrounding cloud
cores (e.g., \citealt{menten1991,sobolev1993}). Class I CH$_3$OH maser
observations at 95\,GHz in DR21 revealed that they are located at the
interface of the molecular outflow and the surrounding ambient gas
\citep{plambeck1990}. Similar results were found for Class I CH$_3$OH
maser emission near 25\,GHz in the Orion-KL region, where
\citet{johnston1992} suggest that the Class I maser emission is formed
behind shock fronts between the outflows and the dense cloud gas. 

\begin{figure}
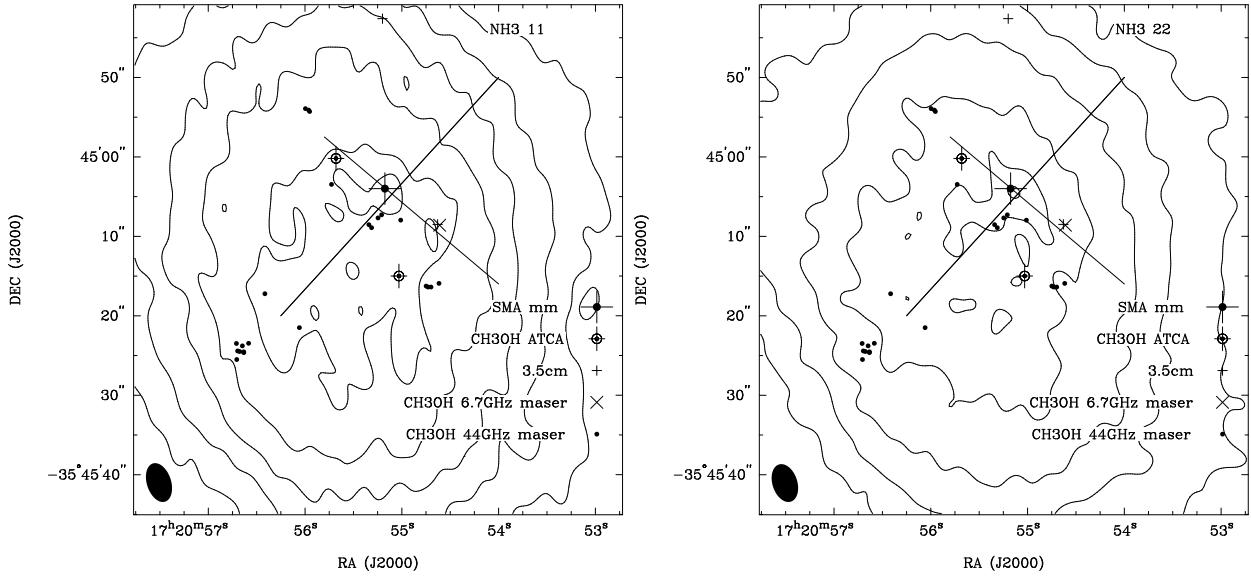

\includegraphics[angle=-90,width=8.2cm]{f2a.ps}
\includegraphics[angle=-90,width=8.2cm]{f2b.ps}
\caption{Integrated main hyperfine component NH$_3$ (1,1) and (2,2)
line emission toward NGC6334\,I(N) (left and right, respectively). The
two lines outline the directions of the two observed outflows in the
region (north-east south west Hunter et al.~in prep., south-east north
west \citealt{megeath1999}). The contour levels are
58(58)348\,mJy\,beam$^{-1}$ ($3\sigma\sim 9$\,mJy\,beam$^{-1}$) and 33(33)198\,mJy\,beam$^{-1}$
($3\sigma\sim 6$\,mJy\,beam$^{-1}$) for the NH$_3$(1,1) and (2,2),
respectively. The symbols are marked at the bottom-right of each
panel. The synthesized beams are shown at the bottom-left of each
panel as well.}
\label{ngc6334in_images}
\end{figure}

In the case of NGC6334\,I(N), the north-eastern 25\,GHz maser feature
(the $-3$\,km\,s$^{-1}$ component) is obviously spatially associated
with the jet-like SiO(5--4) outflow (Hunter et al.~in prep., e.g.,
Fig.~\ref{ngc6334in_images}), and thus fits well into the
shock-excitement picture. The southern 25\,GHz maser (the
$-5$\,km\,s$^{-1}$ component) is less clearly associated with one of
the two outflow axes. However, the north-west south-eastern outflow of
\citet{megeath1999} is significantly less collimated than the newly
detected SiO jet of Hunter et al.~(in prep.), and therefore it is
plausible that the Class I maser features near the southern
$-5$\,km\,s$^{-1}$ maser peak (i.e., 25 and 44\,GHz masers) are
associated with the interface of the broader outflow cone and the
ambient cloud. Comparing the overall spatial distribution of the 25
and 44\,GHz Class I CH$_3$OH maser emission with the present outflow
morphologies (see Fig.~\ref{ngc6334in_images}) all maser features
appear to be associated with the outflows and are likely caused by the
shock-interactions of the outflows and the ambient gas.

The Class I CH$_3$OH masers in NGC6334\,I(N) thus show the same
behavior as masers of this kind in many other regions: First,
individual narrow-line maser components are spread over areas of
several tenths of a parsec, offset from any possible exciting
source. Additional evidence places them in the interaction zones of
outflows and dense ambient gas.  Second, apart from perhaps Orion-KL,
masers in the 44\,GHz transition are always stronger than those 
in any of the 25\,GHz lines, and frequently the strongest of all Class
I CH$_3$OH masers.  This can be understood from the (collisional)
pumping mechanism of Class I methanol masers, which produces efficient
masing in the 25\,GHz lines over just the range of physical parameters
one may expect in the scenario described above (Leurini, Menten, \&\
Schilke 2005, in prep.). This means that possible 25 GHz emission from
many of the spots from where 44 GHz emission arises may be too weak to be
detectable.

\section{Summary}

The NH$_3$(1,1) \& (2,2) and CH$_3$OH observations of the two massive
twin cores NGC6334\,I \& I(N) reveal significantly different emission
signatures between both regions. While NGC6334\,I shows strong,
compact thermal emission in the observed NH$_3$ and CH$_3$OH lines
adjacent to the UCH{\sc ii} region and coincident with mm continuum
peaks observed by Hunter et al.~(in prep.), there is no compact
thermal line emission toward NGC6334\,I(N). Here, we find that the
NH$_3$ emission is distributed over a broad area of size approximately
$1'$. No thermal CH$_3$OH emission is found toward NGC6334\,I(N) but
we do detect two Class I CH$_3$OH maser peaks with peak brightness
temperatures of $\sim 7000$\,K. The two maser peaks may be spatially
associated with 44\,GHz Class I CH$_3$OH maser emission and are likely
produced in the interfaces between the molecular outflows and the
ambient gas.

NH$_3$ peak brightness temperatures $> 70$\,K toward the mm-peak positions
in NGC6334\,I \& I(N) indicate kinetic temperatures of the same
order. However, the NH$_3$(1,1) \& (2,2) lines are only suitable as a
good thermometer below 50\,K and thus we cannot derive more than a
lower limit for the temperatures in these regions. Directly at the
western edge of the UCH{\sc ii} region we find a hot spot where the
line ratios between the NH$_3$(2,2) and (1,1) outermost satellite
hyperfine components are surprisingly high around 4, indicating even
higher temperatures. Since this position is not associated with any mm
continuum peak, the heating is likely caused by the UCH{\sc ii}
region. NH$_3$ emission is also detected toward the outflow in
NGC6334\,I, and we can estimate a rotational temperature of approximately
$19$\,K there.

Toward the UCH{\sc ii} region in NGC6334\,I we observe CH$_3$OH and
NH$_3$ absorption. The absorption features are blue-shifted with
regard to the emission features close-by. This signature is consistent
with expanding molecular gas around the UCH{\sc ii} region.  We do not
find additional strong kinematic effects imposed from the UCH{\sc ii}
region on the adjacent molecular cores.

\acknowledgments{We like to thank Tyler Bourke and Paul Ho for helpful
ATCA and NH$_3$ discussion during the analysis and interpretation of
the data. We are also grateful to an anonymous referee who commented
carefully and detailed on the draft, and helped improving it.
H.B. acknowledges financial support by the Emmy-Noether-Program of the
Deutsche Forschungsgemeinschaft (DFG, grant BE2578/1), and S.T. is
grateful to the Alexander von Humboldt foundation for a Feodor Lynen
research fellowship.}

\clearpage

\clearpage

\clearpage

\clearpage

\clearpage

\clearpage

\clearpage

\clearpage

\clearpage

\clearpage

\begin{figure}
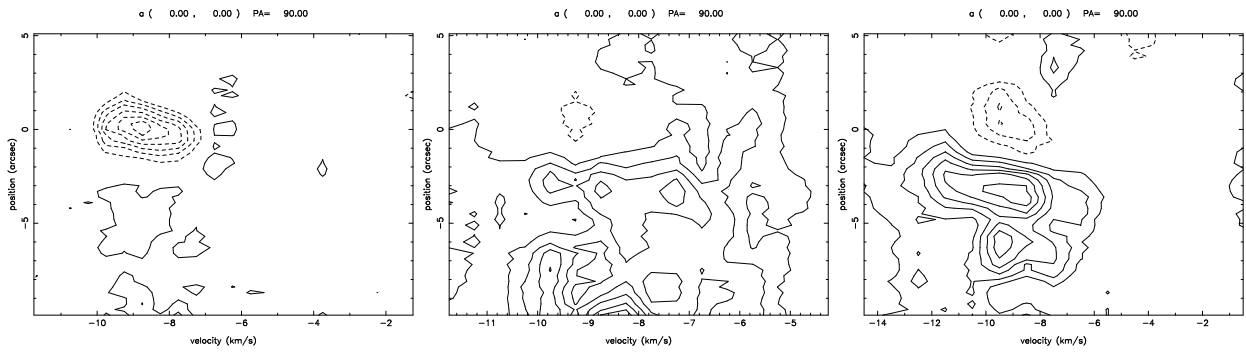

\includegraphics[angle=-90,width=5.4cm]{f10a.ps}
\includegraphics[angle=-90,width=5.4cm]{f10b.ps}
\includegraphics[angle=-90,width=5.4cm]{f10c.ps}
\caption{Position-velocity cuts in the R.A. direction at
$\Delta$Dec.$=0''$ for the \methone~line (left) and the two
NH$_3$(1,1) and (2,2) main components (middle and right) in
NGC6334\,I. Positive emission is in full contours and negative
emission in dashed contours. The contour levels are in $\pm 10$, $\pm
10$ and $\pm 9.5$\,mJy\,beam$^{-1}$ steps for the left, middle and right
panel, respectively.}
\label{pv_absorption}
\end{figure}

\clearpage

\clearpage

\clearpage

\clearpage

\begin{deluxetable}{lrr}
\tablecaption{Phase reference centers\label{center}}
\tablewidth{0pt}
\tablehead{
\colhead{Source} & \colhead{R.A.} & \colhead{Dec.} \\
& \colhead{[J2000]} & \colhead{[J2000]} \\
}
\startdata
NGC6334\,I    & 17:20:53.44 & -35:47:02.2 \\
NGC6334\,I(N) & 17:20:54.63 & -35:45:08.9 \\
\enddata
\end{deluxetable}

\begin{deluxetable}{lrr}
\tablecaption{CH$_3$OH Class I maser positions\label{maserpos}}
\tablewidth{0pt}
\tablehead{
\colhead{Velocity} & \colhead{R.A.} & \colhead{Dec.} \\
\colhead{[km\,s$^{-1}$]} & \colhead{[J2000]} & \colhead{[J2000]} \\
}
\startdata
$-3$ & 17:20:55.68 & -35:45:00.2 \\
$-5$ & 17:20:55.03 & -35:45:15.0
\enddata
\tablecomments{The estimated absolute positional accuracy is $\sim 0.2''$.}
\end{deluxetable}

\begin{deluxetable}{lrrr}
\tablecaption{NH$_3$(1,1) satellite hyperfine components \label{linewidths}}
\tablewidth{0pt}
\tablehead{
\colhead{Source} & \colhead{Offset} & \colhead{$\Delta v$} & \colhead{E} \\
& \colhead{[$''$]} & \colhead{[km\,s$^{-1}$]} & \colhead{[erg]}\\
}
\startdata
NGC6334\,I & -1.0,4.5 & 5.1 & $1.2\times 10^{46}$ \\
NGC6334\,I & -3.0,3.5 & 4.1 & $4.5\times 10^{45}$\\
NGC6334\,I(N) & 6.5,5.0 & 4.5 & $4.8\times 10^{46}$
\enddata
\tablecomments{The line-widths are derived from Gaussian fits to the outermost NH$_3$(1,1) hyperfine component at most negative velocities (see Figures \ref{ngc6334i_nh3_spectra} \& \ref{ngc6334in_nh3_spectra}).}
\end{deluxetable}

\end{document}